\documentclass[12pt]{article}
\usepackage{enumerate}
\usepackage{amsfonts}
\usepackage{amsmath}
\usepackage{amssymb}
\usepackage{amsthm}
\usepackage{latexsym}
\usepackage{color}

\def\H{\mathcal{H}}

\def\S{\mathfrak{S}}
\def\C{\mathfrak{C}}
\def\T{\mathfrak{T}}
\def\B{\mathfrak{B}}

\newcommand{\supp}{\mathrm{supp}}
\newcommand{\rank}{\mathrm{rank}}
\newcommand{\id}{\mathrm{Id}}
\newcommand{\Tr}{\mathrm{Tr}}

\newcommand{\shs}{\hspace{1pt}}
\newcounter{defin}  \newcounter{lemma}  \newcounter{theorem}
\newcounter{property} \newcounter{corol}  \newcounter{remark} \newcounter{example}
\newenvironment{lemma}{\par\refstepcounter{lemma}     \textbf{Lemma \thelemma.} }{\rm\par}
\newenvironment{theorem}{\par\refstepcounter{theorem}     \textbf{Theorem \thetheorem.}\ }{\rm\par}
\newenvironment{property}{\par\refstepcounter{property}     \textbf{Proposition \theproperty.}\ }{\rm\par}
\newenvironment{corollary}{\par\refstepcounter{corol}     \textbf{Corollary \thecorol.} }{\rm\par}

\newenvironment{remark}{\par\refstepcounter{remark}     \textbf{Remark \theremark.}}{\rm\par}
\newenvironment{example}{\par\refstepcounter{example}     \textbf{Example \theexample.}}{\rm\par}

\begin{document}

\title{Advanced Alicki-Fannes-Winter method for energy-constrained quantum systems and its use}
%\title{Universal tight continuity bounds for characteristics of energy-constrained quantum systems and channels}
\author{M.E.~Shirokov \\
Steklov Mathematical Institute, Moscow, Russia}
\date{}
\maketitle
%\vspace{50pt}
\begin{abstract}
We describe an universal method for quantitative continuity analysis of entropic characteristics of energy-constrained quantum systems and channels.
It gives asymptotically tight continuity bounds for basic  characteristics of quantum systems of wide class (including multi-mode quantum oscillators) and channels between such systems  under the energy constraint.

The main application of the proposed  method is the advanced version of the uniform finite-dimensional approximation theorem for  basic capacities of energy-constrained quantum channels.
\end{abstract}

\tableofcontents

\section{Introduction}

Quantitative continuity analysis of characteristics of quantum systems and channels is important for different tasks of quantum information theory.\footnote{A very noncomplete list of \emph{this and previous years}  papers which results are based on using this or that continuity bound is the following \cite{Datta, VJ, AL-1,AL-2,AL-3,AL-4,AL-10,AL-5,Pir,Pir+,AL-6,AL-7,AL-8,AL-11,AL-9}.} This is confirmed by a number of works devoted to this question \cite{A&F, Aud, A&E, A&E+, Fannes, H&D, L&S, W&R, CHI, AFM, CID, SR&H, Wilde++, W-CB}.

The first result in this direction is the famous Fannes' continuity bound (estimate for variation) for the von Neumann entropy in a finite dimensional quantum system used essentialy in the proofs of many theorems in quantum information theory \cite{Fannes}. An optimized version of Fannes's continuity bound obtained by Audenaert in \cite{Aud} states that
\begin{equation*}%\label{Aud-CB}
 |H(\rho)-H(\sigma)|\leq \varepsilon \ln (d-1)+h_2(\varepsilon),\quad  \varepsilon=\textstyle\frac{1}{2}\|\rho-\sigma\|_1,
\end{equation*}
for any states $\rho$ and $\sigma$ in a $d$-dimensional Hilbert space provided that $\varepsilon\leq  1-1/d$. \smallskip

Another important result is the Alicki-Fannes continuity bound for the quantum conditional entropy  obtained in \cite{A&F}  by using the elegant geometric method. This continuity bound is also used essentially in applications, in particular, it allows to prove uniform continuity of the squashed entanglement $E_{\mathrm{sq}}$ (one of the basic entanglement measures)  on the set of all states of a finite-dimensional bipartite system (in fact, it is the necessity to prove the continuity of $E_{\mathrm{sq}}$ that motivated the research by Alicki and Fannes, c.f.\cite{C&W}). The method
 used in \cite{A&F} was then improved by different authors (c.f.\cite{M&H,SR&H}). The optimal version of this method was proposed and used by Winter in \cite{W-CB} to obtain
tight continuity bound for the  quantum conditional entropy and for the relative entropy of entanglement.
In fact, this method (in what follows we will call it the Alicki-Fannes-Winter method, briefly, the AFW-method) is quite universal, it gives uniform continuity bound  for any bounded function $f$ on the set $\S(\H)$ of quantum states which is \emph{locally almost affine}  in the following sense
\begin{equation}\label{a-a-f}
-a(p)\leq f(p\rho+(1-p)\sigma)-p f(\rho)-(1-p)f(\sigma)\leq b(p),
\end{equation}
for arbitrary states $\rho$ and $\sigma$ in $\S(\H)$ and any $p\in(0,1)$, where $a(p)$ and $b(p)$ are  nonnegative functions on $(0,1)$ vanishing as $p\rightarrow0^+$. In quantum information theory the following classes of functions satisfying this condition are widely used:
\begin{itemize}
  \item real linear combinations of marginal entropies of a state of a composite quantum system and their compositions with quantum channels and operations;
  \item basic characteristics of a quantum channels and operations (the output entropy, the entropy exchange, the mutual and coherent informations);
  \item relative entropy distances from a state to a given convex set of states.
\end{itemize}

In particular, the AFW-method shows that any locally almost affine bounded function on $\S(\H)$ is uniformly continuous on $\S(\H)$.

The AFW-method can be used regardless of the dimension of the underlying Hilbert space $\H$ under the condition that $f$ is a bounded function on
the whole set of states. But in  analysis of infinite-dimensional quantum systems we often deal with functions which are bounded only on the sets of states with bounded energy, i.e. states $\rho$ satisfying the inequality
\begin{equation}\label{b-e-ineq}
\Tr H\rho\leq E,
\end{equation}
where $H$ is a positive operator --  Hamiltonian of a quantum system associated with the space $\H$ \cite{H-SCI,H-c-w-c,CID, Wilde+,W-CB}.\smallskip

Winter was the first who proposed a way for quantitative continuity analysis of characteristics of infinite-dimensional quantum systems under the energy constraint (\ref{b-e-ineq}). In \cite{W-CB}  he obtained asymptotically tight continuity bounds for the von Neumann entropy and for the quantum conditional entropy under the energy constraint by using the two-step  approach based on the AFW-method combined with finite-dimensional approximation
of states with bounded energy. Winter's approach was used in \cite{CHI} to obtain asymptotically tight continuity bounds for the quantum conditional mutual information under the energy constraint on one subsystem.\smallskip

Application of Winter's method to any function $f$ possessing property (\ref{a-a-f}) on the set of states with bounded energy is limited by the approximation step, since it requires special estimates depending on this function.
An attempt to obtain an universal continuity bound for functions on the set of states with bounded energy was made in \cite{AFM},
where the method using initial purification of states followed by the standard AFW-technique is proposed. This method allows to
obtain continuity bounds for various characteristics of quantum systems and channels under different forms of energy constraints \cite{AFM,CID}. It plays
a central  role in the proof of the uniform finite-dimensional approximation theorem for basic capacities of infinite-dimensional energy-constrained quantum channels \cite{UFA}.

The main drawback of the universal continuity bound proposed in \cite{AFM} is its non-accuracy: the main term of the
upper bound for $|f(\rho)-f(\sigma)|$ depends on $\sqrt{\varepsilon}$, where $\varepsilon=\frac{1}{2}\|\rho-\sigma\|_1$. This is
a corollary of the initial purification of the states $\rho$ and $\sigma$.

The main aim of this paper is to propose universal continuity bounds for any function $f$ possessing property (\ref{a-a-f}) on the set of states with bounded energy
which would be tight or close-to-tight (asymptotically, for large energy bound). Our method is  close to Winter's two-step approach mentioned before but uses completely different approximation step based on the special property of quantum states with bounded energy  stated in Lemma 3 in \cite{CID}. This approximation step exploits only  property (\ref{a-a-f}) of a function $f$ (via the AFW-technique) and does not require anything else.
As a result the whole two-step method becomes quite universal and accurate (Theorem \ref{SCB-1}).\smallskip

The arguments used in the construction of universal continuity bound can be applied to show existence of
appropriate infinite-dimensional extensions  for characteristics of finite-dimensional $n$-partite quantum systems and channels (Theorem \ref{SCB-2}).

\smallskip

The paper is organized as  follows.

The essence of the proposed method is described in Section 3.1 in full generality (Theorems 1 and 2). Then, in Section 3.2, the specification of this method to the case of a multi-mode quantum oscillator is considered (Corollary 1).\smallskip

In Section 4 we apply the general results of Section 3 to concrete characteristics of quantum system and channels. In particular, we essentially improve the continuity bounds for the quantum mutual information, the coherent information and the output Holevo quantity of a quantum channel under the input energy constraint previously obtained in \cite{AFM,CID}.\smallskip

The main application of the proposed method is the advanced version of the uniform finite-dimensional approximation theorem for  basic capacities of energy-constrained quantum channels  presented in Section 5. Efficiency of the obtained estimates of the $\varepsilon$-sufficient input dimensions for all the basic capacities is confirmed by numerical calculations with the one-mode quantum oscillator in the role of the input system.

\section{Preliminaries}

\subsection{Basic notations}

Let $\mathcal{H}$ be a separable infinite-dimensional Hilbert space,
$\mathfrak{B}(\mathcal{H})$ the algebra of all bounded operators on $\mathcal{H}$ with the operator norm $\|\cdot\|$ and $\mathfrak{T}( \mathcal{H})$ the
Banach space of all trace-class
operators on $\mathcal{H}$  with the trace norm $\|\!\cdot\!\|_1$. Let
$\mathfrak{S}(\mathcal{H})$ be  the set of quantum states (positive operators
in $\mathfrak{T}(\mathcal{H})$ with unit trace) \cite{H-SCI,N&Ch,Wilde}.

Denote by $I_{\mathcal{H}}$ the identity operator on a Hilbert space
$\mathcal{H}$ and by $\id_{\mathcal{\H}}$ the identity
transformation of the Banach space $\mathfrak{T}(\mathcal{H})$.\smallskip

The \emph{von Neumann entropy} of a quantum state
$\rho \in \mathfrak{S}(\H)$ is  defined by the formula
$H(\rho)=\operatorname{Tr}\eta(\rho)$, where  $\eta(x)=-x\ln x$ for $x>0$
and $\eta(0)=0$. It is a concave lower semicontinuous function on the set~$\mathfrak{S}(\H)$ taking values in~$[0,+\infty]$ \cite{H-SCI,L-2,W}.
The von Neumann entropy satisfies the inequality
\begin{equation}\label{w-k-ineq}
H(p\rho+(1-p)\sigma)\leq pH(\rho)+(1-p)H(\sigma)+h_2(p)
\end{equation}
valid for any states  $\rho$ and $\sigma$ in $\S(\H)$ and $p\in(0,1)$, where $\,h_2(p)=\eta(p)+\eta(1-p)\,$ is the binary entropy \cite{N&Ch,Wilde}.\smallskip

The \emph{quantum relative entropy} for two states $\rho$ and
$\sigma$ in $\mathfrak{S}(\mathcal{H})$ is defined as
$$
H(\rho\,\|\shs\sigma)=\sum\langle
i|\,\rho\ln\rho-\rho\ln\sigma\,|i\rangle,
$$
where $\{|i\rangle\}$ is the orthonormal basis of
eigenvectors of the state $\rho$ and it is assumed that
$H(\rho\,\|\sigma)=+\infty$ if $\,\mathrm{supp}\rho\shs$ is not
contained in $\shs\mathrm{supp}\shs\sigma$ \cite{H-SCI,L-2}.\footnote{The support $\mathrm{supp}\rho$ of a state $\rho$ is the closed subspace spanned by the eigenvectors of $\rho$ corresponding to its positive eigenvalues.}\smallskip

The \emph{quantum conditional entropy}
\begin{equation*}%\label{c-e-d}
H(X|Y)_{\rho}=H(\rho)-H(\rho_{\shs Y})
\end{equation*}
of a  state $\rho$ in $\S(\H_{XY})$ with finite marginal entropies is essentially used in analysis of quantum systems \cite{H-SCI,Wilde}. The quantum conditional entropy
can be extended to the set of all states $\rho$ with finite $H(\rho_X)$ by the formula
\begin{equation}\label{ce-ext}
H(X|Y)=H(\rho_{X})-H(\rho\shs\Vert\shs\rho_{X}\otimes
\rho_{Y})
\end{equation}
proposed in \cite{Kuz}.  This extension  possesses all basic properties of the quantum conditional entropy valid in finite dimensions \cite{Kuz,CMI}. In particular, it is concave and satisfies the inequality
\begin{equation*}%\label{CE-ULB}
|H(X|Y)_{\rho}|\leq H(\rho_X)
\end{equation*}
for arbitrary state $\rho$ in $\S(\H_{XY})$ with finite $H(\rho_X)$. \smallskip

The \emph{quantum mutual information} of a state $\,\rho\,$ of a
bipartite quantum system $XY$ is defined as
\begin{equation}\label{mi-d}
I(X\!:\!Y)_{\rho}=H(\rho\shs\Vert\shs\rho_{X}\otimes
\rho_{\shs Y})=H(\rho_{X})+H(\rho_{\shs Y})-H(\rho),
\end{equation}
where the second formula is valid if $\,H(\rho)\,$ is finite \cite{L-mi}.

The \emph{quantum conditional mutual information (QCMI)} of a state $\rho$ of a
tripartite finite-dimensional system $XYZ$ is defined as
\begin{equation}\label{cmi-d}
    I(X\!:\!Y|Z)_{\rho}\doteq
    H(\rho_{XZ})+H(\rho_{\shs YZ})-H(\rho)-H(\rho_{Z}).
\end{equation}
This quantity plays important role in quantum
information theory \cite{D&J,Wilde}, its nonnegativity is a basic result well known as \emph{strong subadditivity
of von Neumann entropy} \cite{Ruskai}. If system $Z$ is trivial then (\ref{cmi-d}) coincides with (\ref{mi-d}).\smallskip

In infinite dimensions formula (\ref{cmi-d}) may contain the uncertainty
$"\infty-\infty"$. Nevertheless the
conditional mutual information can be defined for any state
$\rho$ in $\S(\H_{XYZ})$ by the expression
\begin{equation}\label{cmi-e+}
I(X\!:\!Y|Z)_{\rho}=\sup_{P_X}\left[\shs I(X\!:\!YZ)_{Q_X\rho
Q_X}-I(X\!:\!Z)_{Q_X\rho Q_X}\shs\right],\;\; Q_X=P_X\otimes I_{YZ},\!
\end{equation}
where the supremum is over all finite rank projectors
$P_X\in\B(\H_X)$ and it is assumed that $I(X\!:\!Y')_{Q_X\rho
Q_X}=\lambda I(X\!:\!Y')_{\lambda^{-1} Q_X\rho
Q_X}$, where $\lambda=\Tr\shs Q_X\rho$ \cite{CMI}.\smallskip

Expression (\ref{cmi-e+}) defines the  lower semicontinuous nonnegative function on the set
$\S(\H_{XYZ})$ coinciding with the r.h.s. of (\ref{cmi-d}) for any state $\rho$ at which it is well defined and  possessing all basic properties of the quantum conditional mutual information valid in finite dimensions \cite[Th.2]{CMI}. In particular,
\begin{equation}\label{CMI-UB}
I(X\!:\!Y|Z)_{\rho}\leq 2\min\left\{H(\rho_X),H(\rho_{\shs Y}),H(\rho_{XZ}),H(\rho_{\shs YZ}) \right\}
\end{equation}
for arbitrary state $\rho$ in $\S(\H_{XYZ})$ and
\begin{equation}\label{CMI-M}
I(XR\!:\!Y|Z)_{\rho}\geq I(X\!:\!Y|Z)_{\rho}
\end{equation}
for arbitrary state $\rho$ in $\S(\H_{XYZR})$, where $R$ is any quantum system.

\subsection{The set of quantum states with bounded energy}\label{sec:22}

Let $H_A$ be a positive (semi-definite) operator on a Hilbert space $\mathcal{H}_A$ and $\mathcal{D}(H_A)$ its domain.\footnote{Sometimes we will deal with positive operators which are not densely defined.}  We will assume that
\begin{equation}\label{H-as}
\mathrm{Tr} H_A\rho=
\left\{\begin{array}{l}
        \sup_n\mathrm{Tr} P_n H_A\rho\;\; \textrm{if}\;\;  \supp\rho\subseteq {\rm cl}(\mathcal{D}(H_A))\\
        +\infty\;\;\textrm{otherwise}
        \end{array}\right.
\end{equation}
for any positive operator $\rho\in\T(\H_A)$, where ${\rm cl}(\mathcal{D}(H_A))$ is the closure of $\mathcal{D}(H_A)$
and $P_n$ is the spectral projector of $H_A$ corresponding to the interval $[0,n]$. For any bounded function $f$ we will assume that $\Tr f(H_A)$ is a trace over the Hilbert space ${\rm cl}(\mathcal{D}(H_A))$.\smallskip

Let $E_0$ be the infimum of the spectrum of $H_A$ and $E\geq E_0$. Then
$$
\mathfrak{C}_{H_A,E}=\left\{\rho\in\mathfrak{S}(\mathcal{H}_A)\,|\,\mathrm{Tr} H_A\rho\leq E\right\}
$$
is a closed convex subset of $\mathfrak{S}(\mathcal{H}_A)$. If
$H_A$ is treated as  Hamiltonian of a quantum system $A$ then
$\mathfrak{C}_{H_A,E}$  is the set of states with the mean energy not exceeding $E$.\smallskip

It is well known that the von Neumann entropy is continuous on the set $\mathfrak{C}_{H_A,E}$ for any $E> E_0$ if (and only if) the Hamiltonian  $H_A$ satisfies  the condition
\begin{equation}\label{H-cond}
  \mathrm{Tr}\, e^{-\lambda H_{A}}<+\infty\quad\textrm{for all}\;\lambda>0
\end{equation}
and that the maximal value of the entropy on this set is achieved at the \emph{Gibbs state} $\gamma_A(E)\doteq e^{-\lambda(E) H_A}/\mathrm{Tr} e^{-\lambda(E) H_A}$, where the parameter $\lambda(E)$ is determined by the equality $\mathrm{Tr} H_A e^{-\lambda(E) H_A}=E\mathrm{Tr} e^{-\lambda(E) H_A}$ \cite{W}. Condition (\ref{H-cond}) implies that $H_A$ is an unbounded operator having  discrete spectrum of finite multiplicity. It can be represented as follows
\begin{equation}\label{H-rep}
H_A=\sum_{k=0}^{+\infty} E_k |\tau_k\rangle\langle\tau_k|,
\end{equation}
where
$\left\{\tau_k\right\}_{k=0}^{+\infty}$ is the orthonormal
basis of eigenvectors of $H_A$ corresponding to the nondecreasing sequence $\left\{\smash{E_k}\right\}_{k=0}^{+\infty}$ of eigenvalues
tending to $+\infty$.

We will use the function
\begin{equation}\label{F-def}
F_{H_A}(E)\doteq\sup_{\rho\in\mathfrak{C}_{H_{\!A},E}}H(\rho)=H(\gamma_A(E)).
\end{equation}
It is easy to show that $F_{H_A}$ is a strictly increasing concave function on $[E_0,+\infty)$ such that $F_{H_A}(E_0)=\ln m(E_0)$, where $m(E_0)$ is the multiplicity of $E_0$ \cite{W-CB}.

In this paper we will assume that the Hamiltonian $H_A$ satisfies the condition
\begin{equation}\label{H-cond+}
  \lim_{\lambda\rightarrow0^+}\left[\mathrm{Tr}\, e^{-\lambda H_A}\right]^{\lambda}=1,
\end{equation}
which is slightly stronger than condition (\ref{H-cond}). In terms of the sequence $\{E_k\}$ of eigenvalues of $H_A$
condition (\ref{H-cond}) means that $\lim_{k\rightarrow\infty}E_k/\ln k=+\infty$, while (\ref{H-cond+}) is valid  if $\;\liminf_{k\rightarrow\infty} E_k/\ln^q k>0\,$ for some $\,q>2$ \cite[Proposition 1]{AFM}.
By Lemma 1 in \cite{AFM} condition (\ref{H-cond+}) holds if and only if
\begin{equation}\label{H-cond++}
  F_{H_A}(E)=o\shs(\sqrt{E})\quad\textrm{as}\quad E\rightarrow+\infty,
\end{equation}
while condition (\ref{H-cond}) is equivalent to  $\,F_{H_A}(E)=o\shs(E)\,$ as $\,E\rightarrow+\infty$.
It is essential that condition (\ref{H-cond+})  holds for the Hamiltonians of many real quantum systems \cite{Datta,AFM}.\footnote{Theorem 3 in \cite{Datta} shows that $F_{H_A}(E)=O\shs(\ln E)$ as $E\rightarrow+\infty$ provided that condition (\ref{BD-cond}) holds.}

The function
\begin{equation}\label{F-bar}
 \bar{F}_{H_A}(E)=F_{H_A}(E+E_0)=H(\gamma_A(E+E_0))
\end{equation}
is concave and nondecreasing on $[0,+\infty)$. Let $\hat{F}_{H_A}$ be a continuous function
on $[0,+\infty)$ such that
\begin{equation}\label{F-cond-1}
\hat{F}_{H_A}(E)\geq \bar{F}_{H_A}(E)\quad \forall E>0,\quad \hat{F}_{H_A}(E)=o\shs(\sqrt{E})\quad\textrm{as}\quad E\rightarrow+\infty
\end{equation}
and
\begin{equation}\label{F-cond-2}
\hat{F}_{H_A}(E_1)<\hat{F}_{H_A}(E_2),\quad\hat{F}_{H_A}(E_1)/\sqrt{E_1}\geq \hat{F}_{H_A}(E_2)/\sqrt{E_2}
\end{equation}
for any $E_2>E_1>0$. Sometimes we will additionally assume that
\begin{equation}\label{F-cond-3}
\hat{F}_{H_A}(E)=\bar{F}_{H_A}(E)(1+o(1))\quad\textrm{as}\quad E\to+\infty.
\end{equation}
By property (\ref{H-cond++}) the role of $\hat{F}_{H_A}$ can be played by the function $\bar{F}_{H_A}$
provided that the function  $E\mapsto\bar{F}_{H_A}(E)/\sqrt{E}$ is nonincreasing. In general case the existence of a function $\hat{F}_{H_A}$ with the required properties is established in the following\smallskip

\begin{property}\label{add-l}
A) \emph{If the Hamiltonian $H_A$ satisfies condition (\ref{H-cond+}) then
\begin{equation*}%\label{m-fun}
\hat{F}^{*}_{H_A}(E)\doteq \sqrt{E}\sup_{E'\geq E}\bar{F}_{H_A}(E')/\sqrt{E'}
\end{equation*}
is the minimal function satisfying all the conditions in (\ref{F-cond-1}) and (\ref{F-cond-2}).}

B) \emph{Let
\begin{equation*}
\!N_{\shs\uparrow}[H_A](E)\doteq \sum_{k,j: E_k+E_j\leq E} E_k^2\quad \textrm{and}\quad N_{\downarrow}[H_A](E)\doteq \sum_{k,j: E_k+E_j\leq E} E_kE_j
\end{equation*}
for any $E>E_0$. If
\begin{equation}\label{BD-cond}
\exists \lim_{E\rightarrow+\infty}N_{\shs\uparrow}[H_A](E)/N_{\downarrow}[H_A](E)=a>1
\end{equation}
then
\begin{itemize}
  \item  there is $E_*$ such that the function $E\mapsto\bar{F}_{H_A}(E)/\sqrt{E}$ is nonincreasing for all $E\geq E_*$ and hence $\hat{F}^{*}_{H_A}(E)=\bar{F}_{H_A}(E)$ for all $E\geq E_*$;
  \item $\hat{F}^{*}_{H_A}(E)=(a-1)^{-1} (\ln E)(1+o(1))$ as $E\rightarrow+\infty$.
\end{itemize}}
\end{property}

\emph{Proof.} Part A of the proposition is proved immediately.

By noting that $F_{H_A}(E)=\lambda(E)E+\ln \Tr e^{-\lambda(E)H_A}$ for any $E>E_0$, where $\lambda(E)$
is  defined after (\ref{H-cond}), it is easy to show that
$$
\frac{d}{dE} \left(F_{H_A}(E)/\sqrt{E}\right)=\left(\lambda(E)E-\ln \Tr e^{-\lambda(E)H_A}\right)/(2E\sqrt{E}\,)
$$ 
for any $E>E_0$.
So, part B follows from Theorem 3 in \cite{Datta}. $\square$\smallskip

\textbf{Note:} The condition of part B of Proposition \ref{add-l} is valid for the Hamiltonians
of many real quantum systems \cite{Datta}.\smallskip

Practically, it is convenient to use functions $\hat{F}_{H_A}$ defined by simple formulae. The example of
such function $\hat{F}_{H_A}$ satisfying all the conditions in (\ref{F-cond-1}),(\ref{F-cond-2}) and (\ref{F-cond-3}) in the case when $A$ is a multimode quantum oscillator is considered in Section 3.2.\smallskip

Let $\hat{F}^{-1}_{H_A}$ be the inverse function to the function $\hat{F}_{H_A}$ defined on $[\hat{F}_{H_A}(0), +\infty)$ taking values in $[0, +\infty)$. Let $d_0$ be the minimal natural number such that $\ln d_0>\hat{F}_{H_A}(0)$. The following lemma plays a basic role in this paper.\smallskip

\begin{lemma}\label{cl} \emph{Let $\,\bar{E}\doteq E-E_0\geq 0$, $\,d\geq d_0\,$ and  $\;\gamma(d)\doteq \hat{F}_{H_A}^{-1}(\ln d)$. Let $B$ be any system. If $\,\bar{E}\leq\gamma(d)$ then for any state $\rho$ in $\,\mathfrak{S}(\mathcal{H}_{AB})$ such that $\,\rank\shs \rho_A>d$ and $\mathrm{Tr} H_A\rho_A\leq E$ there exist states $\varrho$, $\sigma_1$ and $\sigma_2$ in $\,\mathfrak{S}(\mathcal{H}_{AB})$ and a number $t\in(0,1]$ such that
$\,\rank\shs \varrho_A\leq d$, $\,\mathrm{Tr} H_A\varrho_A\leq E$, $\,\textstyle\frac{1}{2}\|\rho-\varrho\|_1\leq t \leq \displaystyle\sqrt{\bar{E}/\gamma(d)}$, $\,\mathrm{Tr} H_A[\sigma_k]_A\leq E_0+\bar{E}/t^2$, $k=1,2$, and}
$$
\frac{1}{1+t}\,\rho+\frac{t}{1+t}\,\sigma_1=\frac{1}{1+t}\,\varrho+\frac{t}{1+t}\,\sigma_2.
$$
\end{lemma}

\emph{Proof.} Take a pure state $\hat{\rho}$ in $\S(\H_{ABR})$ such that
$\hat{\rho}_{AB}=\rho$. By Lemma 3 in \cite{CID} there exists\footnote{It is easy to see that the assertion of Lemma 3 in \cite{CID} remains valid with the function $\bar{F}_{H_A}$ replaced by its upper bound $\hat{F}_{H_A}$.} a pure state $\hat{\varrho}$ in $\S(\H_{ABR})$ such that
$\,\rank\shs \hat{\varrho}_A\leq d$, $\,\mathrm{Tr}H_A \hat{\varrho}_A\leq E$, $\,\textstyle\frac{1}{2}\|\hat{\rho}-\hat{\varrho}\|_1\leq \displaystyle\sqrt{\bar{E}/\gamma(d)}\,$ and
$$
\|\hat{\rho}-\hat{\varrho}\|_1\mathrm{Tr}\bar{H}_A\left[[\hat{\rho}-\hat{\varrho}\shs]_{-}\right]_A\leq 2\bar{E},\quad \|\hat{\rho}-\hat{\varrho}\|_1\mathrm{Tr}\bar{H}_A\left[[\hat{\rho}-\hat{\varrho}\shs]_{+}\right]_A\leq 2\bar{E}.
$$
where $[\hat{\rho}-\hat{\varrho}\shs]_-$ and $\,[\hat{\rho}-\hat{\varrho}\shs]_+$  are, respectively, the negative and positive parts of the Hermitian operator $\,\hat{\rho}-\hat{\varrho}\,$ and $\,\bar{H}_A=H_A-E_0I_{A}$.

It is easy to see that  $\hat{\sigma}_1=t^{-1}[\hat{\rho}-\hat{\varrho}]_{-}$ and $\hat{\sigma}_2=t^{-1}[\hat{\rho}-\hat{\varrho}]_{+}$,
where $t=\frac{1}{2}\|\hat{\rho}-\hat{\varrho}\|_1$, are states in $\S(\H_{ABR})$. Then
$$
\frac{1}{1+t}\,\hat{\rho}+\frac{t}{1+t}\,\hat{\sigma}_1=\frac{1}{1+t}\,\hat{\varrho}+\frac{t}{1+t}\,\hat{\sigma}_2.
$$
Hence, the states $\,\varrho=\Tr_R\hat{\varrho}\,$ and $\,\sigma_k=\Tr_R\shs\hat{\sigma}_k$, $k=1,2$, with the above defined parameter $t$ have the required properties, since $\textstyle\frac{1}{2}\|\rho-\varrho\|_1\leq t$
by monotonicity of the trace norm under a partial trace. $\square$

\section{The main results}\label{sec:3}

\subsection{General case}\label{sec:31}

Many important characteristics of states of a $n$-partite
finite-dimensional quantum system $\widehat{X}=X_{1}...X_{n}$ have a form of a function $f$ on the set $\S(\H_{\widehat{X}})$
satisfying the inequalities
\begin{equation}\label{F-p-1}
-a_f h_2(p)\leq f(p\rho+(1-p)\sigma)-p f(\rho)-(1-p)f(\sigma)\leq b_fh_2(p)
\end{equation}
for any states $\rho$  and $\sigma$ in $\S(\H_{\widehat{X}})$ and any $p\in[0,1]$, where $h_2$ is the binary entropy (defined after (\ref{w-k-ineq})) and $\,a_f\,b_f\in \mathbb{R}_+$,
and the inequalities
\begin{equation}\label{F-p-2}
-c^-_f H(\rho_{A})\leq f(\rho)\leq c^+_fH(\rho_{A}),
\end{equation}
for any state $\rho$ in $\S(\H_{\widehat{X}})$, where $A$ is some subsystem of $\widehat{X}$ and $c^-_f,c^+_f\in \mathbb{R}_+$. Examples of characteristics satisfying (\ref{F-p-1}) and (\ref{F-p-2}) are presented in Section 4.

The AFW method (proposed in the optimal form in \cite{W-CB} and
described in a full generality in the proof of Proposition 1 in \cite{CMI}) allows to show that
\begin{equation}\label{fcb}
 |f(\rho)-f(\sigma)|\leq C\varepsilon\ln d+Dg(\varepsilon),\quad d=\dim\H_A,
\end{equation}
for any states $\rho$ and $\sigma$ in $\S(\H_{\widehat{X}})$ such that $\;\frac{1}{2}\|\shs\rho-\sigma\|_1\leq\varepsilon$, where $C=c_f^{+}+c_f^{-}$, $D=a_f+b_f\,$ and
\begin{equation}\label{g-fun}
  g(x)\!\doteq\!(1+x)h_2\!\left(\frac{x}{1+x}\right)=(x+1)\ln(x+1)-x\ln x.
\end{equation}

Assume now that $\widehat{X}=X_{1}...X_{n}$ is a $n$-partite
infinite-dimensional quantum system and $f$ is a function well defined on the set
of all states $\rho$ in $\S(\H_{\widehat{X}})$ with finite energy of $\rho_A$ satisfying conditions (\ref{F-p-1}) and (\ref{F-p-2}) on this set. If the Hamiltonian $H_A$ of the system $A$ satisfies condition (\ref{H-cond+})
then the modification of the AFW method proposed in \cite{AFM} allows to obtain the following continuity bound for this function
\begin{equation}\label{SBC-ineq-}
    |f(\rho)-f(\sigma)|\leq C\sqrt{2\varepsilon}\bar{F}_{H_A}\!\left(\bar{E}/\varepsilon\right)+Dg(\sqrt{2\varepsilon}),\quad \bar{E}=E-E_0,
\end{equation}
which holds for any states $\rho$ and $\sigma$  in $\S(\H_{\widehat{X}})$ such that $\,\Tr H_A\rho_{A},\Tr H_A\sigma_{A}\leq E$ and $\;\frac{1}{2}\|\shs\rho-\sigma\|_1\leq\varepsilon$, where
$\bar{F}_{H_A}$ is the function defined in (\ref{F-bar}) and $E_0$ is the minimal eigenvalue of $H_A$. The r.h.s. of (\ref{SBC-ineq-}) tends to zero as $\,\varepsilon\rightarrow0\,$ due to the property (\ref{H-cond++}) equivalent to (\ref{H-cond+}).
Continuity bound (\ref{SBC-ineq-}) is obtained by purification of the states $\rho$ and $\sigma$ followed by the standard AFW technique based on the property (\ref{F-p-1}) and the inequalities
\begin{equation}\label{F-p-2+}
-c^-_f F_{H_A}(E)\leq f(\rho)\leq c^+_f F_{H_A}(E),
\end{equation}
valid for any state $\rho$ in $\S(\H_{\widehat{X}})$ such that $\Tr H_A\rho_A\leq E$, which follow from (\ref{F-p-2}).
The main drawback of continuity bound (\ref{SBC-ineq-}) is its nonaccuracy for small $\varepsilon$ related to its dependance on $\sqrt{\varepsilon}$ (this is a corollary of the
initial purification of $\rho$ and $\sigma$).

In the following theorem we present more accurate continuity bound for a function $f$ on the sets of states in $\S(\H_{\widehat{X}})$ with finite energy of $\rho_A$ satisfying conditions (\ref{F-p-1}) and (\ref{F-p-2}). We will assume that
$H_A$ is a positive operator on $\H_A$ satisfying condition (\ref{H-cond+}) with the minimal eigenvalue $E_0\geq 0$,  $\hat{F}_{H_A}$ is any continuous function on $\mathbb{R}_+$ satisfying conditions (\ref{F-cond-1}) and  (\ref{F-cond-2}),  $d_0$ is the minimal natural number such that  $\,\ln d_0>\hat{F}_{H_A}(0)\,$ and $\,\gamma(d)=\hat{F}^{-1}_{H_A}(\ln d)\,$ for any $d\geq d_0$.\footnote{The function $\hat{F}^*_{H_A}$ defined in Proposition \ref{add-l} can be used in the role of $\hat{F}_{H_A}$.}\smallskip

\begin{theorem}\label{SCB-1} \emph{Let $f$ be a function on the set $\,\{\shs\rho\in\S(\H_{\widehat{X}})\shs|\,\Tr H_A\rho_{A}<+\infty\shs\}$  satisfying inequalities  (\ref{F-p-1}) and (\ref{F-p-2}). Let $\,\bar{E}\doteq E-E_0>0$, $\varepsilon>0$ and $\,T=(1/\varepsilon)\min\{1, \sqrt{\bar{E}/\gamma(d_0)}\}$. Then
\begin{equation}\label{SBC-ineq}
    |f(\rho)-f(\sigma)|\leq C\varepsilon(1+4t)\!\left(\widehat{F}_{H_{\!A}}\!\!\left[\!\frac{\bar{E}}{(\varepsilon t)^2}\!\right]+\mathrm{\Delta}\right)+D(2g(\varepsilon t)+g(\varepsilon(1+2t)))\!
\end{equation}
for arbitrary states $\rho$ and $\sigma$ in $\S(\H_{\widehat{X}})$ s.t. $\,\Tr H_A\rho_{A},\Tr H_A\sigma_{A}\leq E$ and $\;\frac{1}{2}\|\shs\rho-\sigma\|_1\leq\varepsilon$ and any $\,t\in(0,T]$, where $C=c_f^{+}+c_f^{-}$,  $D=a_f+b_f$ and $\mathrm{\Delta}=1/d_0+\ln2$.}
\smallskip

\emph{If conditions (\ref{F-cond-3}) and (\ref{BD-cond}) hold~\footnote{By Proposition \ref{add-l} this holds, in particular, if  $\hat{F}_{H_A}=\hat{F}^*_{H_A}$.} then the r.h.s. of  (\ref{SBC-ineq}) can be written as
$$
C\varepsilon(1+4t)\!\left(\ln\!\left[\!\frac{\bar{E}}{(\varepsilon t)^2}\!\right]\frac{1+o(1)}{a-1}+\mathrm{\Delta}\right)+D(2g(\varepsilon t)+g(\varepsilon(1+2t))),\quad\varepsilon\rightarrow0^+.
$$
If, in addition, both estimates in (\ref{F-p-2+}) are asymptotically tight for large $E$ in the following sense
\begin{equation}\label{at}
\!\lim_{E\rightarrow+\infty} \left[\frac{\inf_{\rho\in\C_{H_{\!A}\!,\!E}} f(\rho)}{F_{H_A}(E)}+c^-_f\right]=\lim_{E\rightarrow+\infty}\left[c^+_f-\frac{\sup_{\rho\in\C_{H_{\!A}\!,\!E}}f(\rho)}{F_{H_A}(E)}\right]=0,
\end{equation}
where $\C_{H_A,E}=\{\shs\rho\in\S(\H_{\widehat{X}})\shs|\,\Tr H_A\rho_{A}\leq E\shs\}$ and $F_{H_A}$ is the function defined in (\ref{F-def}), then continuity bound (\ref{SBC-ineq}) with optimal $\,t$ is  asymptotically tight for large $E$.\footnote{A continuity bound $\;\displaystyle\sup_{x,y\in S_a}|f(x)-f(y)|\leq B_a(x,y)\;$ depending on a parameter $\,a\,$ is called \emph{asymptotically tight} for large $\,a\,$ if $\;\displaystyle\limsup_{a\rightarrow+\infty}\sup_{x,y\in S_a}\frac{|f(x)-f(y)|}{B_a(x,y)}=1$.}}
\end{theorem}\smallskip

\begin{remark}\label{inc-p}
Since the function $\hat{F}_{H_A}$ satisfies condition (\ref{F-cond-1}) and (\ref{F-cond-2}), the r.h.s. of (\ref{SBC-ineq}) (denoted by $\mathbb{CB}_{\shs t}(\bar{E},\varepsilon\,|\,C,D)$ in what follows) is a nondecreasing function of $\varepsilon$ and $\bar{E}$ tending to zero as $\,\varepsilon\rightarrow0^+$ for any given $\bar{E}$, $C$, $D$ and $\,t\in(0,T]$.
\end{remark}\smallskip

\begin{remark}\label{t-r}
 The "free" parameter $\,t\,$ can be used to optimize continuity bound (\ref{SBC-ineq}) for given values of $E$ and $\varepsilon$.
\end{remark}\smallskip

\emph{Proof.} Let $B=\widehat{X}\setminus A$.  By Lemma \ref{cl} for any $d>d_0$ such that $\bar{E}\leq\gamma(d)$ there exist states $\varrho$, $\varsigma$, $\alpha_k$,  $\beta_k$,  $k=1,2$, in $\,\mathfrak{S}(\mathcal{H}_{AB})$
and numbers $p,q\leq \displaystyle\sqrt{\bar{E}/\gamma(d)}$ such that
$\,\rank\shs \varrho_A,\,\rank\shs \varsigma_A\leq d$, $\,\mathrm{Tr}H_A \varrho_A, \mathrm{Tr}H_A \varsigma_A\leq E$, $\,\textstyle\frac{1}{2}\|\rho-\varrho\|_1\leq p$,
$\,\textstyle\frac{1}{2}\|\sigma-\varsigma\|_1\leq q$,
$\,\mathrm{Tr} \bar{H}_A[\alpha_k]_A\leq \bar{E}/p^2$, $\,\mathrm{Tr} \bar{H}_A[\beta_k]_A\leq \bar{E}/q^2$, $k=1,2$, and
\begin{equation}\label{2-r}
(1-p')\rho+p'\alpha_1=(1-p')\varrho+p'\alpha_2,\quad(1-q')\sigma+q'\beta_1
=(1-q')\varsigma+q'\beta_2,
\end{equation}
where $\bar{H}_A=H_A-E_0I_A$, $\,p'=\frac{p}{1+p}\,$ and $\,q'=\frac{q}{1+q}$. If $\,\rank\shs \rho_A\leq d$ we assume that $\varrho=\rho$ and do not introduce the states $\alpha_k$. Similar assumption holds if $\,\rank\shs \sigma_A\leq d$.

The function $f$ is defined on all the states $\varrho$, $\varsigma$, $\alpha_1$, $\alpha_2$, $\beta_1$, $\beta_2$, since the function $\Tr H_A(\cdot)$ (defined in (\ref{H-as})) is finite at their marginal states corresponding to the subsystem $A$.

By using the first relation in (\ref{2-r}) and inequality (\ref{F-p-1}) it is easy to show that
$$
(1-p')(f(\rho)-f(\varrho))\leq p' (f(\alpha_2)-f(\alpha_1))+(a_f+b_f) h_2(p')
$$
and
$$
(1-p')(f(\varrho)-f(\rho))\leq p' (f(\alpha_1)-f(\alpha_2))+(a_f+b_f) h_2(p').
$$
These inequalities imply that
\begin{equation}\label{one}
|f(\varrho)-f(\rho)|\leq p|f(\alpha_2)-f(\alpha_1)|+(a_f+b_f) g(p).
\end{equation}
Similarly, by using the second relation in (\ref{2-r}) and inequality (\ref{F-p-1}) we obtain
\begin{equation}\label{two}
|f(\varsigma)-f(\sigma)|\leq q|f(\beta_2)-f(\beta_1)|+(a_f+b_f) g(q).
\end{equation}

Since $\,\mathrm{Tr} \bar{H}_A[\alpha_k]_A\leq \bar{E}/p^2$ and  $\,\mathrm{Tr} \bar{H}_A[\beta_k]_A\leq \bar{E}/q^2$ , $k=1,2$, it follows from
(\ref{F-p-2+}) that
\begin{equation}\label{one+}
|f(\alpha_2)-f(\alpha_1)|\leq(c_f^{+}+c_f^{-})\widehat{F}_{H_A}\!\left(\bar{E}/p^2\right)
\end{equation}
and
\begin{equation}\label{two+}
|f(\beta_2)-f(\beta_1)|\leq(c_f^{+}+c_f^{-})\widehat{F}_{H_A}\!\left(\bar{E}/q^2\right).
\end{equation}
Since $p,q\leq y\doteq\displaystyle\sqrt{\bar{E}/\gamma(d)}\,$ and the function $\,E\mapsto\widehat{F}_{H_A}(E)/\sqrt{E}\,$ is non-increasing, we have
$$
x\widehat{F}_{H_A}\!\left(\bar{E}/x^2\right)\leq y\widehat{F}_{H_A}\!\left(\bar{E}/y^2\right)=\sqrt{\bar{E}/\gamma(d)}\,\widehat{F}_{H_A}\!\left(\gamma(d)\right)=
\sqrt{\bar{E}/\gamma(d)}\ln d,
$$
$x=p,q$, where the last equality follows from the definition of $\gamma(d)$.

Thus, it follows from (\ref{one})-(\ref{two+}) and the monotonicity of the function $g(x)$ that
\begin{equation}\label{three+}
|f(\varrho)-f(\rho)|,|f(\varsigma)-f(\sigma)|\leq C\shs\sqrt{\bar{E}/\gamma(d)}\ln d+Dg\!\left(\sqrt{\bar{E}/\gamma(d)}\right),
\end{equation}
where $C=c_f^{+}+c_f^{-}$ and $D=a_f+b_f$.

Since $\,\rank\shs \varrho_A\leq d$ and $\,\rank\shs \varsigma_A\leq d$, the supports of both states
$\varrho_A$ and $\varsigma_A$ are contained in some $2d$-dimensional subspace of $\H_A$. By the triangle inequality
we have
$$
\|\varrho-\varsigma\|_1\leq \|\varrho-\rho\|_1+\|\varsigma-\sigma\|_1+\|\rho-\sigma\|_1\leq 2\varepsilon+4\sqrt{\bar{E}/\gamma(d)}.
$$
So, by using the standard AFW method one can show that\footnote{It is easy to see that all the states used in this method are contained in the domain of $f$.}
\begin{equation}\label{fcb-c}
 |f(\varrho)-f(\varsigma)|\leq \left(2\sqrt{\bar{E}/\gamma(d)}+\varepsilon\right) C\ln (2d)+Dg\!\left(2\sqrt{\bar{E}/\gamma(d)}+\varepsilon\right).
\end{equation}

It follows from (\ref{three+}) and (\ref{fcb-c}) that
\begin{equation}\label{m-cb}
\!\begin{array}{rl}
|f(\rho)-f(\sigma)|\,\leq &  \displaystyle \left(4\sqrt{\bar{E}/\gamma(d)}+\varepsilon\right) C\ln d+\left(2\sqrt{\bar{E}/\gamma(d)}+\varepsilon\right)C\ln 2\\\\ + & \displaystyle D g\!\left(2\sqrt{\bar{E}/\gamma(d)}+\varepsilon\right)+2D g\!\left(\sqrt{\bar{E}/\gamma(d)}\right).
\end{array}
\end{equation}

If $t\in(0,T]$ then, since the sequence $\gamma(d)$ is increasing, there is a natural number $d_*>d_0$  such that $\gamma(d_*)>\bar{E}/(\varepsilon t)^2\geq \bar{E}$ but $\gamma(d_*-1)\leq \bar{E}/(\varepsilon t)^2$. It follows that
$$
\sqrt{\bar{E}/\gamma(d_*)}\leq \varepsilon t\leq 1\quad \textrm{and} \quad
\ln (d_*-1) = \widehat{F}_{H_A}(\gamma(d_*-1))\leq \widehat{F}_{H_A}(\bar{E}/(\varepsilon t)^2),
$$
where the first condition in (\ref{F-cond-2}) was used.
Since $\ln d_*\leq\ln (d_*-1)+1/(d_*-1)\leq\ln (d_*-1)+1/d_0$, inequality (\ref{m-cb}) with $d=d_*$ implies continuity bound (\ref{SBC-ineq}).\smallskip

If conditions (\ref{F-cond-3}) and (\ref{BD-cond}) hold then it follows from part B of Proposition \ref{add-l}
that $\hat{F}_{H_A}(E)=(a-1)^{-1}\ln(E)(1+o(1))$ as $E\to+\infty$.
This implies the asymptotic representation of the r.h.s. of (\ref{SBC-ineq}) and the following relation
\begin{equation}\label{f-rel}
\!\hat{F}_{H_A}\!\left(\bar{E}/(\varepsilon t)^2\right)=\left(F_{H_A}(E)-2(a-1)^{-1}\ln(\varepsilon t)\right)(1+o(1))\quad\textrm{as }\;E \to +\infty.
\end{equation}

Assume that  both estimates in (\ref{F-p-2+}) are asymptotically tight for large $E$. Then
for any $\delta>0$ there
exists $E_{\delta}>0$ such that for any $E>E_{\delta}$ the set $\C_{H_A,E}$ contains states $\rho$ and
$\sigma$ such that
$|f(\rho)-f(\sigma)|\geq (C-\delta)F_{H_A}(E)$.
Since $\frac{1}{2}\|\rho-\sigma\|_1\leq 1$, it follows that for any $\varepsilon>0$ the set $\C_{H_A,E}$ contains states $\rho_{\varepsilon}$ and
$\sigma_{\varepsilon}$ such that\footnote{This can be shown by using the states $\rho_k=\frac{k}{n}\rho+(1-\frac{k}{n})\sigma$, $k=0,1,..,n$, for sufficiently large $n$.}
\begin{equation}\label{t-p}
\textstyle\frac{1}{2}\|\rho_{\varepsilon}-\sigma_{\varepsilon}\|_1\leq \varepsilon\quad \textrm{and}\quad |f(\rho_{\varepsilon})-f(\sigma_{\varepsilon})|\geq \varepsilon (C-\delta)F_{H_A}(E).
\end{equation}

It follows from (\ref{f-rel}) that the r.h.s. of (\ref{SBC-ineq}) with $t=\varepsilon$ has the form $C\varepsilon F_{H_A}(E)+R(\varepsilon,E)$, where
$R(\varepsilon,E)$ is a finite function such that $R(\varepsilon,E)/\left(\varepsilon F_{H_A}(E)\right)$ tends to zero as $(\varepsilon,E)$
tends to $(0,+\infty)$. So, it is easy to show that (\ref{t-p})
implies the  asymptotical tightness of the continuity bound (\ref{SBC-ineq}) for large $E$. $\square$ \smallskip

In Theorem \ref{SCB-1} it is assumed that the function $f$ is defined on the set of all states $\rho$ in $\S(\H_{\widehat{X}})$, where $\widehat{X}=X_{1}...X_{n}$, such that $\Tr H_A\rho_{A}<+\infty$, but often we deal with functions
originally defined and satisfying the inequalities (\ref{F-p-1}) and  (\ref{F-p-2}) only on the subset
\begin{equation}\label{S-f}
\S_{\rm f}(\H_{\widehat{X}})\doteq\left\{\rho\in\S(\H_{\widehat{X}})\,|\; \rank\rho_{X_i}<+\infty,\;\, i=\overline{1,n}\, \right\}
\end{equation}
of $\S(\H_{\widehat{X}})$. This is the case when we want to construct  a characteristic of a infinite-dimensional $n$-partite quantum system
by using its finite-dimensional version. Thus, the question arises about the extension of a function defined on $\S_{\rm f}(\H_{\widehat{X}})$
to larger subsets of $\S(\H_{\widehat{X}})$, in particular, to the set of all states $\shs\rho\in\S(\H_{\widehat{X}})$ such that $\Tr H_A\rho_{A}<+\infty$.
The technique used in the proof of Theorem \ref{SCB-1} gives a partial solution of this question.

We begin with the following simple but useful observation.\smallskip
\begin{lemma}\label{sl} \emph{Let $f$ be a function on the set $\,\S_{\rm f}(\H_{\widehat{X}})$ (defined in (\ref{S-f})) satisfying inequalities (\ref{F-p-1}) and  (\ref{F-p-2}). Then there exists an extension of
this function to the set $\,\C_{\!A}^*\doteq\{\shs\rho\in\S(\H_{\widehat{X}})\shs|\, \rank\rho_{A}<+\infty \shs\}$
satisfying the same inequalities, which  is uniformly continuous on the set
$\C_{\!A}^k\doteq\{\shs\rho\in\S(\H_{\widehat{X}})\shs|\, \rank\rho_{A}<k  \shs\}$
for any natural $k$. This extension (also denoted by $f$) satisfies on the set $\,\C_A^k$ continuity bound (\ref{fcb}) with $d$ replaced by $2k$.}
\end{lemma}\smallskip

\emph{Proof.} Let $\rho$ and $\sigma$ be any states in $\C_k=\C_A^k\cap\S_{\rm f}(\H_{\widehat{X}})$ for any given $k$.
Since  there is a $2k$-dimensional subspace of $\H_A$ containing the supports of both states $\rho_A$ and $\sigma_A$, the standard AFW technique  shows that inequality (\ref{fcb}) holds for the states $\rho$ and $\sigma$ with $d$ replaced by $2k$. It follows that the function $f$ is uniformly continuous on $\C_k$. Since the set $\C_k$ is dense in $\C_A^k$, the function $f$ has a unique uniformly continuous  extension to the set $\C_A^k$ satisfying continuity bound (\ref{fcb}).

Since the extensions of $f$ to the sets $\C_A^k$ and $\C_A^{l}$ agree with each other for any $k$ and $l$ (this follows from their uniqueness), the function $f$ has an  extension to the set $\C_A^*=\bigcup_{k\in\mathbb{N}}\C_A^k$ with the required properties. $\square$\smallskip

\begin{theorem}\label{SCB-2} \emph{Let $f$ be a function on the set $\,\S_{\rm f}(\H_{\widehat{X}})$ (defined in (\ref{S-f}))  satisfying inequalities (\ref{F-p-1}) and (\ref{F-p-2}). If $H_A$ is a positive operator on $\H_A$ satisfying condition (\ref{H-cond+}) then for any $E\geq E_0$ the
function $f$ has a unique uniformly continuous extension to the set $\C_{H_{\!A},E}=\{\shs\rho\in\S(\H_{\widehat{X}})\shs|\, \Tr H_A\rho_A\leq E \shs\}$
satisfying the same inequalities  and continuity bound (\ref{SBC-ineq}) (obtained by means of any appropriate function $\hat{F}_{H_A}$).\footnote{We do not assume that $H_A$ is a densely defined operator. The quantity $\Tr H_A\rho$ for any $\rho$  in $\S(\H_A)$ is defined according to the rule (\ref{H-as}).}}
\end{theorem}\smallskip

\emph{Proof.} By Lemma \ref{sl} the function $f$ has an extension to the set $\C^*_A$ satisfying inequalities (\ref{F-p-1}) and (\ref{F-p-2}). Let $\hat{F}_{H_A}$ be  a function satisfying conditions (\ref{F-cond-1}) and (\ref{F-cond-2}), for example, the function $\hat{F}^*_{H_A}$ described in Proposition \ref{add-l}. Let $\rho$ and $\sigma$ be any states in
$$
 \C^*_{H_{\!A},E}=\{\shs\rho\in\S(\H_{\widehat{X}})\,|\, \Tr H_A\rho_A\leq E,\, \rank \rho_A<+\infty \shs\}
$$
By repeating the arguments from the proof of Theorem \ref{SCB-1} one can prove inequality (\ref{SBC-ineq}) for these states. It implies that the function $f$ is uniformly continuous on the set $\C^*_{H_A,E}$.
Since  the set $\C^*_{H_A,E}$ is dense in $\C_{H_{\!A},E}$,
the function $f$  has a unique uniformly continuous  extension to the set $\C_{H_{\!A},E}$
satisfying inequalities (\ref{F-p-1}) and (\ref{F-p-2}) with the same parameters and continuity bound (\ref{SBC-ineq}). $\square$\smallskip

\begin{example}\label{privacy}
Consider the function
\begin{equation}\label{P-fun}
P(\rho)=I(B\!:\!R)_{\rho}-I(E\!:\!R)_{\rho}=H(\rho_B)-H(\rho_{BR})-H(\rho_{E})+H(\rho_{ER})
\end{equation}
on the set $\S_{\rm f}(\H_{BER})$, where $B$, $E$ and $R$ are any quantum systems.

Since $P(\rho)=H(R|E)_{\rho}-H(R|B)_{\rho}$ for any state $\rho$ in $\S_{\rm f}(\H_{BER})$, by using concavity of the quantum conditional entropy and inequality (\ref{w-k-ineq}) it is easy to show that the function $P$ satisfies the inequality (\ref{F-p-1}) on the set $\S_{\rm f}(\H_{BER})$ with $a_f=b_f=1$.

The monotonicity of the quantum mutual information and upper bound (\ref{CMI-UB}) with trivial $Z$ imply that
$$
\max\{I(B\!:\!R)_{\rho},I(E\!:\!R)_{\rho}\}\leq I(BE\!:\!R)_{\rho}\leq 2H(\rho_{BE})
$$
for any state $\rho\in\S_{\rm f}(\H_{BER})$. So, the function $P$ satisfies inequality  (\ref{F-p-2}) with $A=BE$ and $c^-_f=c^+_f=2$ on the set $\S_{\rm f}(\H_{BER})$. By Theorem \ref{SCB-2}  for any positive operator $H_{BE}$ on $\H_{BE}$  satisfying  condition
(\ref{H-cond+}) and any $E>E_0$ there exists a unique uniformly continuous extension of the function $P$ to the set
$\{\shs\rho\in\S(\H_{BER})\shs|\,\Tr H_{BE}\rho_{BE}\leq E\shs\}$ satisfying inequality (\ref{F-p-1}) with $a_f=b_f=1$ and
inequality  (\ref{F-p-2}) with $A=BE$ and $c^-_f=c^+_f=2$. Note that the r.h.s. of (\ref{P-fun}) is not well defined on the above set.

The function $P$  will be used in Section 4.5 for deriving continuity bound
for the privacy of energy constrained quantum channels.
\end{example}

\subsection{The case when $A$ is the $\ell$-mode quantum oscillator}\label{sec:32}

Assume now that the system $A$ (involved in (\ref{F-p-2})) is the $\,\ell$-mode quantum oscillator with the frequencies $\,\omega_1,...,\omega_{\ell}\,$. The Hamiltonian of this system has the form
\begin{equation}\label{qos-H}
H_A=\sum_{i=1}^{\ell}\hbar \omega_i a_i^*a_i+E_0 I_A,\quad E_0=\frac{1}{2}\sum_{i=1}^{\ell}\hbar \omega_i,
\end{equation}
where $a_i$ and $a^*_i$ are the annihilation and creation operators of the $i$-th mode \cite{H-SCI}. Note that this
Hamiltonian satisfies condition (\ref{BD-cond}) with $a=1+1/\ell$ \cite{Datta,Datta+}.

It is shown  in \cite[Section III.B]{CHI} that in this case the function $F_{H_A}(E)$ defined in (\ref{F-def}) is bounded above by the function
\begin{equation}\label{F-ub}
F_{\ell,\omega}(E)\doteq \ell\ln \frac{E+E_0}{\ell E_*}+\ell,\quad E_*=\left[\prod_{i=1}^{\ell}\hbar\omega_i\right]^{1/\ell}\!\!,%\vspace{-5pt}
\end{equation}
and that upper bound (\ref{F-ub}) is $\varepsilon$-sharp for large $E$. So, the function
\begin{equation}\label{F-ub+}
\bar{F}_{\ell,\omega}(E)\doteq F_{\ell,\omega}(E+E_0)=\ell\ln \frac{E+2E_0}{\ell E_*}+\ell,%\vspace{-5pt}
\end{equation}
is a  upper bound on the function $\bar{F}_{H_A}(E)\doteq F_{H_A}(E+E_0)$  satisfying all the conditions in (\ref{F-cond-1}),(\ref{F-cond-2}) and (\ref{F-cond-3}).
The second condition in (\ref{F-cond-2}) follows from  Lemma 5 in \cite{CID}. By using the function $\bar{F}_{\ell,\omega}$ in the role of function $\hat{F}_{H_A}$
in Theorem \ref{SCB-1} we obtain the following\smallskip

\begin{corollary}\label{SCB-G-1}
\emph{Let $f$ be a function on the set $\,\{\shs\rho\in\S(\H_{\widehat{X}})\shs|\,\Tr H_A\rho_{A}<+\infty\shs\}$  satisfying (\ref{F-p-1}) and (\ref{F-p-2}),
where $A$ is the $\ell$-mode quantum oscillator with the frequencies $\omega_1,...,\omega_{\ell}$. Let
$E>E_0$, $\varepsilon>0$ and $\,T_*=(1/\varepsilon)\min\{1, \sqrt{\bar{E}/E_0}\}$, where $\,\bar{E}=E-E_0$. Then
\begin{equation}\label{SBC-ineq+}
\begin{array}{c}
\displaystyle |f(\rho)-f(\sigma)|\leq C\varepsilon(1+4t)\left(\ell\ln \frac{\bar{E}/(\varepsilon t)^2+2E_0}{\ell E_*}+\ell+\mathrm{\Delta}^*\!\right)\\\\+D(2g(\varepsilon t)+g(\varepsilon(1+2t)))\\\\
 \leq C\varepsilon(1+4t) \left(F_{\ell,\omega}(E)-2\ell\ln(\varepsilon t)+\mathrm{\Delta}^*\right)+D(2g(\varepsilon t)+g(\varepsilon(1+2t)))
\end{array}\!\!
\end{equation}
for any states $\rho$ and $\sigma$ in $\,\S(\H_{\widehat{X}})$ such that $\,\Tr H_A\rho_{A}\leq E,\,\Tr H_A\sigma_{A}\leq E$ and\break $\;\frac{1}{2}\|\shs\rho-\sigma\|_1\leq\varepsilon$ and any $\,t\in(0,T_*]$, where $C=c^-_f+c^+_f$, $D=a_f+b_f$ and $\,\mathrm{\Delta}^*=e^{-\ell}+\ln2$.}\smallskip

\emph{If both relations in (\ref{at}) hold then continuity bound (\ref{SBC-ineq+}) with optimal $\,t$ is  asymptotically tight for large $E$.}
\end{corollary}\smallskip

\emph{Proof.} All the assertions of the corollary directly follow from Theorem \ref{SCB-1}. It suffices to note that in this case
$d_0$ is the minimal natural number not less than $x^{\ell}$, where $x=2E_0e/(\ell E_*)\geq e$, and hence
$$
\gamma(d_0)\doteq\bar{F}^{-1}_{\ell,\omega}(\ln d_0)=(\ell/e)E_*\sqrt[\ell]{d_0}-2E_0\leq(\ell/e)E_*x\sqrt[\ell]{1+e^{-\ell}}-2E_0\leq E_0.
$$

\section{Applications}\label{sec:4}

\subsection{Basic examples}\label{sec:41}

In this section we apply Theorem \ref{SCB-1} to the basic entropic quantities: the von Neumann entropy, the quantum conditional entropy
and the quantum conditional mutual information (QCMI). Asymptotically tight continuity bounds for the entropy  and for the conditional entropy  under the energy constraint  have been obtained by Winter \cite{W-CB}. Asymptotically tight continuity bounds for the QCMI  under the energy constraint
on one of the subsystems  has been obtained in  \cite{CHI} by using Winter's technique. The aim of this section is to show that our technique
also gives asymptotically tight continuity bounds for these quantities \emph{without any claim of their superiority}. In fact, the
continuity bounds obtained by Winter's technique are slightly better   than
their analogues presented below, in particular, they hold under condition (\ref{H-cond}) (which is weaker than condition (\ref{H-cond+})) any do not require a function $\hat{F}_{H_A}$ with properties (\ref{F-cond-1}) and (\ref{F-cond-2}). The main advantage of the method proposed in this paper is its \emph{universality}, consisting, in particular, in possibility to
obtain continuity bounds under different forms of energy constraint (see Example \ref{e-three} and Remark \ref{m-a} below).

In all the below examples $f$ is a function on the set of states of infinite-dimensional composite
system $XY...$  satisfying  the inequality (\ref{F-p-2}) for a particular subsystem $A$ of $XY...$ and some $c^-_f,c^+_f$ and the inequality (\ref{F-p-1}) for some $\,a_f,\,b_f$ on the sets of all states $\rho$ in $\S(\H_{XY...})$ with finite $\Tr H_A\rho_A$, where $H_A$ is a positive operator on $\H_A$  satisfying  condition (\ref{H-cond+}) with the minimal eigenvalue $E_0$. If $H_A$ is not densely defined on $\H_A$ we define the quantity $\Tr H_A\rho$ for any $\rho\in\S(\H_A)$ according to the rule (\ref{H-as}).  We will assume that $\hat{F}_{H_A}$ is a continuous function on $\mathbb{R}_+$ satisfying conditions (\ref{F-cond-1}) and  (\ref{F-cond-2}). Denote by $\mathbb{CB}_{\shs t}(\bar{E},\varepsilon\,|\,C,D)$ the expression in the r.h.s. of (\ref{SBC-ineq}).\smallskip

\begin{example}\label{e-one}
Let $f(\rho)=H(\rho)$ be the von Neumann entropy of a state $\rho$ of a single quantum system $X$. This function
satisfies  inequality (\ref{F-p-1}) with $a_f=0$, $b_f=1$ and
 inequality (\ref{F-p-2}) with $A=X$ and $c^+_f=1$, $c^-_f=0$.
So, for given $E>E_0$ and $\varepsilon>0\,$ Theorem \ref{SCB-1} implies that
\begin{equation}\label{CB-1}
 |H(\rho)-H(\sigma)|\leq \mathbb{CB}_{\shs t}(\bar{E},\varepsilon\,|\,1,1),\qquad \bar{E}=E-E_0,
\end{equation}
for any states $\rho$ and $\sigma$ in $\S(\H_A)$ such that  $\Tr H_A\rho,\Tr H_A\sigma\leq E$ and\break $\frac{1}{2}\|\shs\rho-\sigma\|_1\leq\varepsilon$
and any $t\in(0,T]$, where $T$
is defined in Theorem \ref{SCB-1}. If conditions (\ref{F-cond-3}) and (\ref{BD-cond}) hold then the last assertion of Theorem \ref{SCB-1} shows that  continuity bound (\ref{CB-1}) with optimal $\,t\,$ is asymptotically tight for large $E$. This is true if $A$ is the $\ell$-mode quantum oscillator (and $\hat{F}_{H_A}=\bar{F}_{\ell,\omega}$). In this case (\ref{CB-1}) holds with $\mathbb{CB}_{\shs t}(\bar{E},\varepsilon\,|\,1,1)$ replaced by
the expression in the r.h.s. of (\ref{SBC-ineq+}) with $C=D=1$ for any $t\in(0,T_*]$, where  $\,T_*=(1/\varepsilon)\min\{1, \sqrt{\bar{E}/E_0}\}$.
\end{example}\smallskip

\begin{example}\label{e-two} Let $f(\rho)$ be the extented  \emph{quantum conditional entropy} $H(X|Y)_{\rho}$ defined in (\ref{ce-ext}).
The extended conditional entropy satisfies inequality (\ref{F-p-1}) with $a_f=0$ and $b_f=1$ and inequality (\ref{F-p-2}) with $A=X$ and $c^-_f=c^+_f=1$ \cite{Kuz,CMI,AFM}. So, for given $E>E_0$ and  $\varepsilon>0\,$ Theorem \ref{SCB-1} implies that

\begin{equation}\label{CB-2}
 |H(A|Y)_{\rho}-H(A|Y)_{\sigma}|\leq \mathbb{CB}_{\shs t}(\bar{E},\varepsilon\,|\,2,1),\qquad \bar{E}=E-E_0,
\end{equation}
for any states $\rho$ and $\sigma$ in $\S(\H_{AY})$ such that  $\Tr H_A\rho_{A},\Tr H_A\sigma_{A}\leq E$ and\break  $\frac{1}{2}\|\shs\rho-\sigma\|_1\leq\varepsilon$
and any $t\in(0,T]$, where $T$ is defined in Theorem \ref{SCB-1}. If conditions (\ref{F-cond-3}) and (\ref{BD-cond}) hold then continuity bound (\ref{CB-2}) with optimal $\,t$ is asymptotically tight for large $E$. This is true if $A$ is the $\ell$-mode quantum oscillator (and $\hat{F}_{H_A}=\bar{F}_{\ell,\omega}$). In this case (\ref{CB-2}) holds with $\mathbb{CB}_{\shs t}(\bar{E},\varepsilon\,|\,2,1)$ replaced by
the expression in the r.h.s. of (\ref{SBC-ineq+}) with $C=2, D=1$ for any $t\in(0,T_*]$, where  $\,T_*=(1/\varepsilon)\min\{1, \sqrt{\bar{E}/E_0}\}$.

The tightness of (\ref{CB-2}) can be easily shown by using the last assertion of Theorem \ref{SCB-1}. Indeed, to prove the first  relation in (\ref{at}) one can take any purification of the Gibbs state $\gamma_A(E)$ in $\S(\H_{AY})$ in the role of $\rho$. To show validity of the second relation in (\ref{at}) one can take $\rho=\gamma_A(E)\otimes \sigma$, where $\sigma$ is any state in $\S(\H_Y)$.
\end{example}\smallskip

\begin{example}\label{e-three} Let $f(\rho)$ be the extended QCMI  $I\,(X\!:\!Y|Z)_{\rho}$
defined by the expression (\ref{cmi-e+}). We will consider it as a function on the set of states of the extended
system $XYZR$ (which coincides with $XYZ$ if the system $R$ is trivial).  The extended QCMI
satisfies inequality (\ref{F-p-1}) with $a_f=1$ and $b_f=1$ \cite{CID}.
It satisfies inequality (\ref{F-p-2}) with any of the subsystems
\begin{equation}\label{B-s}
X,\;\; Y,\;\; XR,\;\; YR,\;\; XZ,\;\; YZ,\;\; XZR,\;\; YZR
\end{equation}
in the role of system $A$ and  $\,c^-_f=0$, $c^+_f=2\,$ in all the cases.
This  follows from the nonnegativity of the QCMI, the upper bounds (\ref{CMI-UB}) and the monotonicity of QCMI expressed by inequality (\ref{CMI-M}).
So, for given $E>E_0$ and $\varepsilon>0\,$ Theorem \ref{SCB-1} implies that
\begin{equation}\label{CB-3}
 |I(X\!:\!Y|Z)_{\rho}-I(X\!:\!Y|Z)_{\sigma}|\leq \mathbb{CB}_{\shs t}(\bar{E},\varepsilon\,|\,2,2),\qquad \bar{E}=E-E_0,
\end{equation}
for any states $\rho$ and $\sigma$ in $\S(\H_{XYZR})$ such that
$\Tr H_A\rho_{A},\Tr H_A\sigma_{A}\leq E$ and $\frac{1}{2}\|\shs\rho-\sigma\|_1\leq\varepsilon$ and any $t\in(0,T]$, where $A$ is one of the subsystems in (\ref{B-s}) and $T$ is defined in Theorem \ref{SCB-1}. If conditions (\ref{F-cond-3}) and (\ref{BD-cond}) hold then continuity bound (\ref{CB-3}) with optimal $\,t$ is asymptotically tight for large $E$ (for any choice of $A$). This is true if $A$ is the \break $\ell$-mode quantum oscillator (and $\hat{F}_{H_A}=\bar{F}_{\ell,\omega}$). In this case (\ref{CB-3}) holds with $\mathbb{CB}_{\shs t}(\bar{E},\varepsilon\,|\,2,2)$ replaced by
the expression in the r.h.s. of (\ref{SBC-ineq+}) with $C=D=2$ for any $t\in(0,T_*]$, where  $\,T_*=(1/\varepsilon)\min\{1, \sqrt{\bar{E}/E_0}\}$.

The tightness of (\ref{CB-3}) can be easily shown by using the last assertion of Theorem \ref{SCB-1}.
Indeed, assume that $A=X$ and that the systems $Z$ and $R$ are trivial, so that $f(\rho)=I(A\!:\!Y)_{\rho}$. Then to show the validity of first and the second relations in (\ref{at}) one can take, respectively,  any product state $\rho_A\otimes\sigma_Y$ in $\S(\H_{AY})$  such that $\Tr H_A\rho_A\leq E$ and any purification of the Gibbs state $\gamma_A(E)$ in $\S(\H_{AY})$ in the role of $\rho$.
\end{example}\smallskip

\begin{remark}\label{m-a}
The possibility to take any of the systems in (\ref{B-s})  in the role of system $A$ in continuity bound (\ref{CB-3}) means
possibility to obtain continuity bound for the QCMI under different forms of energy constraint. It is this feature that
implies efficiency of the proposed technique, it will be used essentially in the following sections.
\end{remark}

\subsection{Tight continuity bound for the  QCMI at the output of a local energy-constrained channel}\label{sec:42}

Let $\mathrm{\Phi}$ be a quantum channel from a system $A$ to a system $B$ (completely positive trace preserving linear map from $\T(\H_A)$ into $\T(\H_B)$), $C$ and $D$ any systems. In this subsection we obtain tight continuity bound for the function
$$\rho \mapsto I(B\!:\!D|C)_{\mathrm{\Phi}\otimes\id_{CD}(\rho)}$$
on $\S(\H_{ACD})$ under the energy constraint on $\rho_A$  which essentially refines the continuity bound for this function obtained  in \cite{CID} by using the method from \cite{AFM}.\footnote{We assume that $I\,(B\!:\!D|C)$ is the extended QCMI
defined by (\ref{cmi-e+}).}

Assume that $H_A$ is the Hamiltonian of a quantum system $A$ with the minimal energy $E_0$ satisfying  condition (\ref{H-cond+})
and $\hat{F}_{H_A}$ is any function on $\mathbb{R}_+$ satisfying conditions (\ref{F-cond-1}) and  (\ref{F-cond-2}). Denote by $\mathbb{CB}_{\shs t}(\bar{E},\varepsilon\,|\,C,D)$ the expression in the r.h.s. of (\ref{SBC-ineq}).
\smallskip

\begin{property}\label{MI-CB} \emph{Let $\,\mathrm{\Phi}:A\rightarrow B$ be a quantum channel, $\,C$ and $\,D$ be any systems, $E>E_0$ and $\,\varepsilon>0$. Then
\begin{equation}\label{CMI-CB+}
|I(B\!:\!D|C)_{\mathrm{\Phi}\otimes\id_{CD}(\rho)}-I(B\!:\!D|C)_{\mathrm{\Phi}\otimes\id_{CD}(\sigma)}|\leq \mathbb{CB}_{\shs t}(\bar{E},\varepsilon\,|\,2,2)
\end{equation}
for any states $\rho$ and $\sigma$ in $\S(\H_{ACD})$ such that $\,\Tr H_A\rho_{A},\Tr H_A\sigma_{A}\leq E$ and  $\;\frac{1}{2}\|\shs\rho-\sigma\|_1\leq\varepsilon$ and any $t\in(0,T]$, where $\bar{E}=E-E_0$ and  $T=T(\bar{E},\varepsilon)$ is defined in Theorem \ref{SCB-1}.}

\emph{If conditions (\ref{F-cond-3}) and (\ref{BD-cond}) hold then continuity bound (\ref{CMI-CB+}) with optimal $\,t$ is asymptotically tight for large $E$.
This is true, in particular, if $A$ is the $\ell$-mode quantum oscillator. In this case (\ref{CMI-CB+}) holds with $\mathbb{CB}_{\shs t}(\bar{E},\varepsilon\,|\,2,2)$ replaced by
the expression in the r.h.s. of (\ref{SBC-ineq+}) with $C=D=2$ for any $t\in(0,T_*]$, where  $\,T_*=(1/\varepsilon)\min\{1, \sqrt{\bar{E}/E_0}\}$.}
\end{property}\smallskip

\emph{Proof.} By the Stinespring theorem a quantum channel $\mathrm{\Phi}:A\rightarrow B$ can be represented as
\begin{equation}\label{St-rep}
\mathrm{\Phi}(\rho)=\Tr_E V_{\mathrm{\Phi}}\rho V_{\mathrm{\Phi}}^*,
\end{equation}
where $V_{\mathrm{\Phi}}$ is an isometry from $\H_A$ into $\H_{BE}$ ($\H_E$ is a separable Hilbert space) \cite{H-SCI}. Continuity bound (\ref{CMI-CB+}) can be obtained from continuity bound (\ref{CB-3}) with $A=XR$, where $X=B$ and $R=E$, by identifying $\H_A$ and $H_A$  with the subspace $V_{\mathrm{\Phi}}\H_A$ of $\H_{BE}$ and the operator $V_{\mathrm{\Phi}}H_AV_{\mathrm{\Phi}}^*$ on $\H_{BE}$ correspondingly.\footnote{Since in general the operator $H_{BE}\doteq V_{\mathrm{\Phi}}H_AV_{\mathrm{\Phi}}^*$ is not densely defined on $\H_{BE}$,
the quantity $\Tr H_{BE}\rho$ for a state $\rho$ in $\S(\H_{BE})$ is defined according to the rule (\ref{H-as}).}

If conditions (\ref{F-cond-3}) and (\ref{BD-cond}) hold then the tightness of continuity bound (\ref{CMI-CB+})
follows from the tightness of continuity bound (\ref{CB-3}) in the case $A=X$. $\square$

\subsection{Tight continuity bounds for the mutual information and coherent information of energy constrained channels}\label{sec:43}

In analysis of information properties of a  channel $\,\mathrm{\Phi}$ between finite-dimensional quantum systems $A$ and $B\,$ the quantities
\begin{equation}\label{mi-def}
 I(\mathrm{\Phi},\rho)=H(\rho)+H(\mathrm{\Phi}(\rho))-H(\mathrm{\Phi},\rho)
\end{equation}
and
\begin{equation}\label{ci-def}
 I_c(\mathrm{\Phi},\rho)=H(\mathrm{\Phi}(\rho))-H(\mathrm{\Phi},\rho),
\end{equation}
where $\rho$ is a state in $\S(\H_A)$ and  $H(\mathrm{\Phi},\rho)$ is the entropy exchange, are widely used. They are called, respectively, the \emph{mutual information} and the \emph{coherent information} of a quantum channel $\mathrm{\Phi}$ at a state $\rho$ \cite{H-SCI,Wilde}.

In infinite dimensions these quantities are well defined for any input state $\rho$ with finite entropy by the expressions
\begin{equation}\label{mi-def+}
 I(\mathrm{\Phi},\rho)=I(B\!:\!R)_{\mathrm{\Phi}\otimes\mathrm{Id}_{R}(\hat{\rho})},
\end{equation}
and
\begin{equation}\label{ci-def+}
 I_c(\mathrm{\Phi},\rho)=I(B\!:\!R)_{\mathrm{\Phi}\otimes\mathrm{Id}_{R}(\hat{\rho})}-H(\rho),
\end{equation}
where $\mathcal{H}_R\cong\mathcal{H}_A$ and $\hat{\rho}\shs$
is a pure state in $\S(\H_{AR})$ such that $\hat{\rho}_{A}=\rho$ \cite{H-Sh-2}.

For any quantum channel $\mathrm{\Phi}:A\rightarrow B$ the inequalities
\begin{equation}\label{mi-b}
0\leq I(\mathrm{\Phi},\rho)\leq 2H(\rho),
\end{equation}
and
\begin{equation}\label{ci-b}
-H(\rho)\leq I_c(\mathrm{\Phi},\rho)\leq H(\rho).
\end{equation}
hold for any state $\rho$ in $\S(\H_A)$ with finite entropy. They follow from the expressions (\ref{mi-def+}) and (\ref{ci-def+}) and the well known
properties of the quantum mutual information.

Continuity  bound for the function $\rho\rightarrow I(\mathrm{\Phi},\rho)$ under the energy constraint on the input states
was obtained in \cite[Corollary 6]{AFM}. It allows to prove uniform continuity  of the function $\rho\rightarrow I(\mathrm{\Phi},\rho)$ on the set of states $\rho$ such that $\Tr H_A\rho\leq E$ for any $E>E_0$ provided that the Hamiltonian $H_A$ satisfies the condition (\ref{H-cond+}).
The main drawback of that continuity  bound is its  nonaccuracy for small $\varepsilon=\frac{1}{2}\|\rho-\sigma\|_1$  connected with its dependance on $\sqrt{\varepsilon}$.

The technique proposed in Section 3 allows to essentially refine the continuity  bound obtained in \cite{AFM}. To apply this technique we will need the following
lemma, in which it is assumed that $I(\mathrm{\Phi},\rho)$ and $I_c(\mathrm{\Phi},\rho)$ are defined by formulae (\ref{mi-def+}) and (\ref{ci-def+}).\smallskip

\begin{lemma}\label{MI-LAA-l} \emph{Let $\mathrm{\Phi}:A\rightarrow B$ be an arbitrary quantum channel.  Then
\begin{equation}\label{MI-LAA}
0\leq I(\mathrm{\Phi},p\rho+(1-p)\sigma)- pI(\mathrm{\Phi},\rho)-(1-p)I(\mathrm{\Phi},\sigma)\leq 2h_2(p)
\end{equation}
and
\begin{equation}\label{CI-LAA}
-h_2(p)\leq I_c(\mathrm{\Phi},p\rho+(1-p)\sigma)- pI_c(\mathrm{\Phi},\rho)-(1-p)I_c(\mathrm{\Phi},\sigma)\leq h_2(p)
\end{equation}
for  any states $\rho$ and $\sigma$ in $\,\S(\H_A)$ with finite entropy and $p\in(0,1)$.}
\end{lemma}\smallskip

\emph{Proof.} The first inequality in (\ref{MI-LAA}) means concavity of the function $\rho\rightarrow I(\mathrm{\Phi},\rho)$
defined in (\ref{mi-def+}). It is proved in \cite{H-Sh-2} (by using the well known concavity of this function in the finite-dimensional case).

All the other inequalities in (\ref{MI-LAA}) and (\ref{CI-LAA}) are easily proved provided that the  channel $\mathrm{\Phi}$ has finite-dimensional output.
Indeed, in this case $H(\mathrm{\Phi}(\rho))$ is finite for any input state $\rho$. So, if $H(\rho)$ is finite then
the entropy exchange $H(\mathrm{\Phi},\rho)$ (coinciding with the output entropy of a complementary channel) is also finite by the triangle
inequality \cite{H-SCI}. Hence, for any state $\rho$ with finite entropy $H(\rho)$ the quantities  $I(\mathrm{\Phi},\rho)$ and $I_c(\mathrm{\Phi},\rho)$ are well defined by formulae (\ref{mi-def}) and (\ref{ci-def}). So, in this case (\ref{CI-LAA}) and the second inequality in (\ref{MI-LAA}) follow from the concavity of the entropy and inequality (\ref{w-k-ineq}).

If $\mathrm{\Phi}$ is an arbitrary quantum channel then there is a sequence $\{\mathrm{\Phi}_n\}$ of channels with finite dimensional output strongly
converging to the channel $\mathrm{\Phi}$, i.e. $\lim_{n\rightarrow+\infty}\mathrm{\Phi}_n(\rho)=\mathrm{\Phi}(\rho)$ for any input state $\rho$. Proposition 10 in \cite{CMI}
implies that
\begin{equation}\label{I-lr}
\lim_{n\rightarrow+\infty}I(\mathrm{\Phi}_n,\rho)=I(\mathrm{\Phi},\rho)\quad\textrm{and}\quad \lim_{n\rightarrow+\infty}I_c(\mathrm{\Phi}_n,\rho)=I_c(\mathrm{\Phi},\rho)
\end{equation}
for any input state $\rho$ with finite entropy. It was mentioned before that (\ref{CI-LAA}) and the second inequality in (\ref{MI-LAA})
hold with $\mathrm{\Phi}=\mathrm{\Phi}_n$ for all $n$. So, it follows from (\ref{I-lr}) that these inequalities hold for the channel $\mathrm{\Phi}$ as well. $\square$\smallskip

Assume that  $H_A$ is the Hamiltonian of a system $A$ with the minimal energy $E_0$ satisfying  condition (\ref{H-cond+})
and $\hat{F}_{H_A}$ is a function on $\mathbb{R}_+$ satisfying conditions (\ref{F-cond-1}) and  (\ref{F-cond-2}). Denote by $\mathbb{CB}_{\shs t}(\bar{E},\varepsilon\,|\,C,D)$ the expression in the r.h.s. of (\ref{SBC-ineq}).
\smallskip

\begin{property}\label{CMI-CB} \emph{Let $\,\mathrm{\Phi}:A\rightarrow B$ be a quantum channel, $E>E_0$ and $\varepsilon>0$. Then
\begin{equation}\label{MI-CB+}
|I(\mathrm{\Phi},\rho)-I(\mathrm{\Phi},\sigma)|\leq \mathbb{CB}_{\shs t}(\bar{E},\varepsilon\,|\,2,2),
\end{equation}
\begin{equation}\label{CI-CB+}
|I_c(\mathrm{\Phi},\rho)-I_c(\mathrm{\Phi},\sigma)|\leq \mathbb{CB}_{\shs t}(\bar{E},\varepsilon\,|\,2,2)
\end{equation}
for any states $\rho$ and $\sigma$ in $\,\S(\H_A)$ such that $\,\Tr H_A\rho,\Tr H_A\sigma\leq E$ and $\;\frac{1}{2}\|\shs\rho-\sigma\|_1\leq\varepsilon$ and any $\,t\in(0,T]$, where $\bar{E}=E-E_0$ and
$T=T(\bar{E},\varepsilon)$ is defined in Theorem  \ref{SCB-1}.}

\emph{If conditions (\ref{F-cond-3}) and (\ref{BD-cond}) hold then continuity bounds (\ref{MI-CB+}) and (\ref{CI-CB+})  with optimal $\,t$ is asymptotically tight for large $E$.
This is true, in particular, if $A$ is the $\ell$-mode quantum oscillator. In this case (\ref{MI-CB+}) and (\ref{CI-CB+}) hold for any $\,t\in(0,T_*]$ with the r.h.s. replaced by
\begin{equation}\label{MI-CB++}
2\varepsilon (1+4t)\left(F_{\ell,\omega}(E)-2\ell\ln(\varepsilon t)+ e^{-\ell}+\ln 2\right)+4g(\varepsilon t)+2g(\varepsilon(1+2t)),
\end{equation}
where $F_{\ell,\omega}$ is the function defined in (\ref{F-ub}) and $\,T_*=(1/\varepsilon)\min\{1, \sqrt{\bar{E}/E_0}\}$.}\smallskip

\emph{If $\,\mathrm{\Phi}$ is  an antidegradable channel (cf.\cite{DC,Wilde}) then  (\ref{MI-CB+}) and (\ref{CI-CB+}) hold with\break
$\mathbb{CB}_{\shs t}(\bar{E},\varepsilon\,|\,1,2)$ in the r.h.s. and the first factor $2$ in (\ref{MI-CB++}) can be removed.}\smallskip

\emph{If $\,\mathrm{\Phi}$ is  degradable  channel (cf.\cite{DC,Wilde}) then  (\ref{CI-CB+}) holds with
$\mathbb{CB}_{\shs t}(\bar{E},\varepsilon\,|\,1,2)$ in the r.h.s. and the first factor $2$ in (\ref{MI-CB++}) can be removed.}

\end{property}\smallskip

\textbf{Note:} The r.h.s. of (\ref{MI-CB+}) and (\ref{CI-CB+}) coincide and do not depend on a channel $\mathrm{\Phi}$. They tend to zero as $\varepsilon\rightarrow0$  for any given $\bar{E}$ and $\shs t\shs$ due to the second condition in (\ref{F-cond-1}) implying uniform continuity of the functions
$\,\rho\mapsto I(\mathrm{\Phi},\rho)$ and $\,\rho\mapsto I_c(\mathrm{\Phi},\rho)$ on the set of states with bounded energy.\smallskip

\emph{Proof.}  Continuity bounds (\ref{MI-CB+}) and (\ref{CI-CB+}) and their specifications for the case
when $A$ is the $\ell$-mode quantum oscillator are derived from  Theorem \ref{SCB-1} and Corollary \ref{SCB-G-1} by using   (\ref{mi-b}), (\ref{ci-b})  and Lemma \ref{MI-LAA-l}.

Assume that conditions (\ref{F-cond-3}) and (\ref{BD-cond}) hold. To show the asymptotic tightness of continuity bound (\ref{MI-CB+})  assume that $\mathrm{\Phi}$ is the noiseless channel, $\rho$ is the Gibbs  state $\gamma_A(E)$ for given $E>E_0$ and  $\sigma$ is any pure state in $\S(\H_A)$ such that $\Tr H_A\sigma\leq E$.
Then
\begin{equation*}%\label{E-ulb}
  I(\mathrm{\Phi},\rho)=2H(\gamma_A(E))=2F_{H_A}(E)\quad \textrm{and} \quad I(\mathrm{\Phi},\sigma)=0.
\end{equation*}
Thus, the asymptotic tightness of continuity bound (\ref{MI-CB+}) follows from the last assertion of Theorem \ref{SCB-1}.

To show the asymptotic tightness of continuity bound (\ref{CI-CB+}) assume that $A$ is the
one mode quantum oscillator with the frequency $\omega$. In this case
\begin{equation*}%\label{G-rep}
\gamma_A(E)=(1-p_E)\sum_{k=0}^{+\infty} p_E^{\shs k} |\eta_k\rangle\langle\eta_k|,
\end{equation*}
where $p_E=(E-E_0)/(E+E_0)$, $E_0=\frac{1}{2}\hbar\omega$, and $\{\eta_k\}_{k=0}^{+\infty}$ is the Fock basis \cite{H-SCI}.

Assume that $E\gg E_0$ and $q=p_{E'}$, where $E'=E-4E_0$. Consider the states
$$
\rho_1=(1-q^2)\sum_{k=1}^{+\infty} q^{2(k-1)} |\eta_{2k}\rangle\langle\eta_{2k}|\quad \textrm{and} \quad
\rho_2=(1-q^2)\sum_{k=1}^{+\infty} q^{2(k-1)} |\eta_{2k-1}\rangle\langle\eta_{2k-1}|
$$
and the channel $\mathrm{\Phi}(\rho)=P\rho P+[\Tr Q\rho]|\eta_0\rangle\langle \eta_0|$, where
$P=\sum_{k=0}^{+\infty}|\eta_{2k}\rangle\langle\eta_{2k}|$ and $Q=I_A-P$.
It is easy to show that $H(\rho_1)=H(\rho_2)\geq H(\gamma_A(E'))-\ln 2$  and that $\Tr H_A\rho_1,\Tr H_A\rho_2\leq E$. So, by direct calculation
we obtain
$$
I_c(\mathrm{\Phi},\rho_1)=H(\rho_1)\geq F_{H_A}(E')-\ln2,\quad   I_c(\mathrm{\Phi},\rho_2)=-H(\rho_2)\leq -F_{H_A}(E')+\ln2.
$$
Since in this case $F_{H_A}(E)=g((E-E_0)/(2E_0))$, these  inequalities imply validity of both limit relations in (\ref{at}) for
the function $\rho\mapsto I_c(\mathrm{\Phi},\rho)$. Thus, the asymptotic tightness of continuity bound (\ref{CI-CB+}) follows from the last assertion of Theorem \ref{SCB-1}.

If $\,\mathrm{\Phi}$ is  an antidegradable channel then the r.h.s. of (\ref{mi-b}) and (\ref{ci-b})
can be replaced, respectively, by $H(\rho)$ and $0$. If $\,\mathrm{\Phi}$ is  a degradable channel then the l.h.s. of  (\ref{ci-b})
can be replaced by $0$.
These observations imply the last assertions of the proposition.
$\square$

\subsection{Close-to-tight continuity bound for the output Holevo quantity
of energy constrained channels}\label{sec:44}

The technic proposed in Section 3 allows to essentially
strengthen Proposition 7 in \cite{CID}.\smallskip

Let $\mathrm{\Phi}:A\rightarrow B$ be an arbitrary quantum channel and $\{p_i,\rho_i\}$ a \emph{discrete ensemble}
of input states -- a finite or
countable collection $\{\rho_{i}\}\subset\S(\H_A)$ with the corresponding probability distribution $\{p_{i}\}$.
The output Holevo quantity of this ensemble under the channel $\mathrm{\Phi}$ is defined as
$$
\chi(\{p_i,\mathrm{\Phi}(\rho_i)\})\doteq \sum_{i} p_i H(\mathrm{\Phi}(\rho_i)\|\mathrm{\Phi}(\bar{\rho}))=H(\mathrm{\Phi}(\bar{\rho}))-\sum_{i} p_i H(\mathrm{\Phi}(\rho_i)),
$$
where $\bar{\rho}=\sum_{i} p_i\rho_i$ is the average state of $\{p_i,\rho_i\}$ and the second formula is valid under the condition $H(\mathrm{\Phi}(\bar{\rho}))<+\infty$. It is an important characteristic related to the classical capacity of a quantum channel \cite{H-SCI,Wilde}.\smallskip

In analysis of continuity of the Holevo quantity we will use three  measures of divergence between ensembles $\mu=\{p_i,\rho_i\}$ and $\nu=\{q_i,\sigma_i\}$ described in detail in \cite{O&C,CHI}.

The quantity
\begin{equation*}%\label{D-0-metric}
D_0(\mu,\nu)\doteq\frac{1}{2}\sum_i\|\shs p_i\rho_i-q_i\sigma_i\|_1
\end{equation*}
is an easily computable  metric on the set of all discrete ensembles of quantum states considered as \emph{ordered} collections of states with the corresponding probability distributions.\smallskip

From the quantum information point of view it is natural to consider an ensemble of quantum states $\{p_i,\rho_i\}$ as a discrete probability measure $\sum_i p_i\delta(\rho_i)$  on the set $\mathfrak{S}(\mathcal{H})$ (where $\delta(\rho)$ is the Dirac measure concentrated at a state $\rho$) rather than ordered (or disordered) collection of states. If we want to identify ensembles corresponding to the same probability measure then we have to use the factorization of $D_0$, i.e. the quantity
 \begin{equation*}%\label{f-metric}
D_*(\mu,\nu)\doteq \inf_{\mu'\in \mathcal{E}(\mu),\nu'\in \mathcal{E}(\nu)}D_0(\mu',\nu')
\end{equation*}
as a measure of divergence between ensembles $\mu=\{p_i,\rho_i\}$ and $\nu=\{q_i,\sigma_i\}$, where $\mathcal{E}(\mu)$ and $\mathcal{E}(\nu)$ are the sets
of all countable ensembles corresponding to the measures $\sum_i p_i\delta(\rho_i)$ and $\sum_i q_i\delta(\sigma_i)$ respectively.  The factor-metric $D_*$ coincides with the EHS-distance $D_{\mathrm{ehs}}$ between ensembles of quantum states proposed by Oreshkov and Calsamiglia in \cite{O&C}. It is obvious that
\begin{equation}\label{d-ineq}
D_*(\mu,\nu)\leq D_0(\mu,\nu)
\end{equation}
for any ensembles $\mu$ and $\nu$.

We will also use the Kantorovich distance
\begin{equation}\label{K-D-d}
D_K(\mu,\nu)=\frac{1}{2}\inf_{\{P_{ij}\}}\sum P_{ij}\|\rho_i-\sigma_j\|_1
\end{equation}
between ensembles $\mu=\{p_i,\rho_i\}$ and $\nu=\{q_i,\sigma_i\}$ of quantum states, where the infimum is taken over all joint probability distributions $\{P_{ij}\}$  such that $\sum_jP_{ij}=p_i$ for all $i$ and $\sum_iP_{ij}=q_j$ for all $j$. It is shown in \cite{O&C} that
\begin{equation}\label{d-ineq+}
  D_*(\mu,\nu)\leq D_K(\mu,\nu)
\end{equation}
for any discrete ensembles $\mu$ and $\nu$.\smallskip

In the study of infinite-dimensional quantum systems and channels  the notion of \textit{generalized (continuous) ensemble} defined as
a Borel probability measure on the set of quantum states is widely used \cite{H-SCI,H-Sh-1}. We denote by $\mathcal{P}(\mathcal{H})$ the set of all Borel probability measures on $\mathfrak{S}(\mathcal{H})$. It contains the  subset $\mathcal{P}_0(\mathcal{H})$ of discrete measures (corresponding to discrete ensembles). The average state of a generalized
ensemble $\mu \in \mathcal{P}(\mathcal{H})$ is defined as the barycenter of the measure
$\mu $, that is
$\bar{\rho}(\mu)=\int_{\mathfrak{S}(\mathcal{H})}\rho \mu (d\rho )$.\smallskip

The Kantorovich distance (\ref{K-D-d}) is extended to generalized ensembles $\mu$ and $\nu$ by the expression
\begin{equation}\label{K-D-c}
D_K(\mu,\nu)=\frac{1}{2}\inf_{\Lambda\in\Pi(\mu,\nu)}\int_{\mathfrak{S}(\mathcal{H})\times\mathfrak{S}(\mathcal{H})}\|\rho-\sigma\|_1\Lambda(d\rho,d\sigma),
\end{equation}
where $\Pi(\mu,\nu)$ is the set of all Borel probability measures on $\mathfrak{S}(\mathcal{H})\times\mathfrak{S}(\mathcal{H})$ with the marginals $\mu$ and $\nu$. Since  $\frac{1}{2}\|\rho-\sigma\|_1\leq 1$ for any states $\rho$ and $\sigma$, the Kantorovich distance (\ref{K-D-c}) generates the weak convergence on the set $\mathcal{P}(\mathcal{H})$  \cite{Bog}.\footnote{A sequence $\{\mu_n\}$ of measures weakly converges to a measure $\mu_0$ if
$\,\lim_{n\rightarrow\infty}\int f(\rho)\mu_n(d\rho)=\int f(\rho)\mu_0(d\rho)\,$
for any continuous bounded function $f$ on $\,\mathfrak{S}(\mathcal{H})$ \cite{Bog}.}\smallskip

For an ensemble $\mu \in \mathcal{P}(\mathcal{H}_{A})$ its image $\mathrm{\Phi}(\mu) $
under a quantum channel $\mathrm{\Phi}:A\rightarrow B\,$ is defined as the
ensemble in $\mathcal{P}(\mathcal{H}_{B})$ corresponding to the measure $\mu
\circ \mathrm{\Phi} ^{-1}$ on $\mathfrak{S}(\mathcal{H}_{B})$, i.e. $\,\mathrm{\Phi} (\mu )[%
\mathfrak{S}_{B}]=\mu[\mathrm{\Phi} ^{-1}(\mathfrak{S}_{B})]\,$ for any Borel subset $%
\mathfrak{S}_{B}$ of $\mathfrak{S}(\mathcal{H}_{B})$, where $\mathrm{\Phi} ^{-1}(%
\mathfrak{S}_{B})$ is the pre-image of $\mathfrak{S}_{B}$ under the map $%
\mathrm{\Phi} $. If $\mu =\{p _{i},\rho _{i}\}$ then  $\mathrm{\Phi} (\mu)=\{p _{i},\mathrm{\Phi}(\rho_{i})\}$.\smallskip

For a given channel $\,\mathrm{\Phi}:A\rightarrow B\,$ the output Holevo quantity of a
generalized ensemble $\mu$ in $\mathcal{P}(\H_A)$ is defined as
\begin{equation*}
\chi(\mathrm{\Phi}(\mu))=\int_{\mathfrak{S}(\mathcal{H})} H(\mathrm{\Phi}(\rho)\shs \|\shs \mathrm{\Phi}(\bar{\rho}(\mu)))\mu (d\rho )=H(\mathrm{\Phi}(\bar{\rho}(\mu
)))-\int_{\mathfrak{S}(\mathcal{H})} H(\mathrm{\Phi}(\rho))\mu (d\rho ),  %\label{chi-q-d+}
\end{equation*}%
where the second formula is valid under the condition $H(\mathrm{\Phi}(\bar{\rho}(\mu)))<+\infty$ \cite{H-Sh-1}.

Assume that $H_A$ is the Hamiltonian of system $A$ with the minimal energy $E_0$ satisfying condition (\ref{H-cond+})
and $\hat{F}_{H_A}$ is a function on $\mathbb{R}_+$ satisfying conditions (\ref{F-cond-1}) and  (\ref{F-cond-2}). Denote by $\mathbb{CB}_{\shs t}(\bar{E},\varepsilon\,|\,C,D)$ the expression in the r.h.s. of (\ref{SBC-ineq}).\smallskip

The following proposition contains continuity bound for the function $\mu\mapsto \chi(\mathrm{\Phi}(\mu))$ under the constraint on the average energy of $\mu$, i.e. under the condition
\begin{equation}\label{av-en}
E(\mu)\doteq \mathrm{Tr} H_A\bar{\rho}(\mu)=\int\mathrm{Tr} H_A\rho\shs\mu(d\rho)\leq E.
\end{equation}

\begin{property}\label{HQ-C-be-1} \emph{Let $\,\mathrm{\Phi}:A\rightarrow B$ be a quantum channel, $E>E_0$ and $\varepsilon>0$. Then
\begin{equation}\label{HQ-CB}
\left|\chi(\mathrm{\Phi}(\mu))-\chi(\mathrm{\Phi}(\nu))\right|\leq \mathbb{CB}_{\shs t}(\bar{E},\varepsilon\,|\,2,2)
\end{equation}
for any ensembles $\shs\mu$ and $\shs\nu$ such that $\,E(\mu),E(\nu)\leq E$ and  $\;D_K(\mu,\nu)\leq\varepsilon$ and any $t\in(0,T]$, where $\bar{E}=E-E_0$ and $\shs T=T(\bar{E},\varepsilon)$ is defined in Theorem \ref{SCB-1}.}

\emph{If $\,\mu\shs$ and $\,\nu\shs$ are discrete ensembles then  the Kantorovich metric $D_K$ can be replaced by any of the metrics  $\,D_0$ and $\,D_*$.}

\emph{If conditions (\ref{F-cond-3}) and (\ref{BD-cond}) hold then continuity bound (\ref{HQ-CB}) with optimal $\,t$ is close-to-tight for large $E$ up to factor $2$ in the main term. This is true, in particular, if $A$ is the $\ell$-mode quantum oscillator. In this case (\ref{HQ-CB}) holds with $\mathbb{CB}_{\shs t}(\bar{E},\varepsilon\,|\,2,2)$ replaced by
the expression in the r.h.s. of (\ref{SBC-ineq+}) with $C=D=2$ for any $t\in(0,T_*]$, where  $\,T_*=(1/\varepsilon)\min\{1, \sqrt{\bar{E}/E_0}\}$.}
\end{property}\smallskip

\textbf{Note:} The r.h.s. of (\ref{HQ-CB}) does not depend on a channel $\mathrm{\Phi}$. It tends to zero as $\varepsilon\rightarrow0$
for any given $\bar{E}$ and $\shs t \shs $ due to the second condition in (\ref{F-cond-1}).\smallskip

\emph{Proof.} The arguments from the proof of Corollary 7 in \cite{AFM} with the use of Proposition \ref{MI-CB} in Section 4.2 (instead of Proposition 5 in \cite{AFM})
implies validity of the inequality (\ref{HQ-CB}) any discrete ensembles $\mu$ and $\nu$ such that $\,E(\mu),E(\nu)\leq E$ and  $D_*(\mu,\nu)\leq\varepsilon$. It follows from (\ref{d-ineq}) and (\ref{d-ineq+}) that this inequality holds for any $\;\varepsilon\geq D(\mu,\nu)$, where $D$ is either $D_0$ or $D_K$.

If $\mu$ and $\nu$ are arbitrary generalized ensembles such that $\,E(\mu),E(\nu)\leq E$ and  $D_K(\mu,\nu)\leq\varepsilon$ then inequality (\ref{HQ-CB}) can be proved by repeating the arguments from the proof of Proposition 7 in \cite{CID} based on approximation of $\mu$ and $\nu$ by weakly converging sequences of discrete ensembles.

Assume that conditions (\ref{F-cond-3}) and (\ref{BD-cond}) hold. To show that in this case continuity bound (\ref{HQ-CB}) with optimal $\,t$ is close-to-tight for large $E$ one should assume that $\mathrm{\Phi}$ is the ideal channel,  take an ensemble $\mu$
consisting of the single Gibbs state $\gamma_A(E)$  and ensemble $\nu$ of pure states with the average state $\gamma_A(E)$ and to repeat the arguments from the proof of the last assertion of Theorem \ref{SCB-1}.

The specification of (\ref{HQ-CB}) to the case when $A$ is a multi-mode oscillator follows from Corollary \ref{SCB-G-1}. $\square$

\subsection{Continuity bound for the privacy
of energy constrained channels}\label{sec:45}

In this subsection we will use the notion of a (generalized) ensemble of quantum states
and different measures of divergence between such ensembles briefly described in the previous subsection.
Let $\mathrm{\Phi}:A\rightarrow B$ be an arbitrary quantum channel with the Stinespring represententation (\ref{St-rep}). Then the channel
\begin{equation}\label{c-ch}
\widehat{\mathrm{\Phi}}(\rho)=\Tr_B V_{\mathrm{\Phi}}\rho V_{\mathrm{\Phi}}^*,
\end{equation}
from the system $A$ to the environment $E$ is called complementary to the channel \cite{H-SCI}.

The privacy of a quantum channel $\mathrm{\Phi}$ at a discrete or continuous ensemble $\mu$
of input states is defined as
$$
\pi_{\mathrm{\Phi}}(\mu)=\chi(\mathrm{\Phi}(\mu))-\chi(\widehat{\mathrm{\Phi}}(\mu)),
$$
provided that this difference is well defined \cite{H-SCI}.\footnote{The term "privacy" is used in the literature in different senses.} Despite the fact that
complementary channel (\ref{c-ch}) to the channel $\mathrm{\Phi}$ depends on the Stinespring representation (\ref{St-rep}) of $\mathrm{\Phi}$,
its output Holevo quantity $\chi(\widehat{\mathrm{\Phi}}(\mu))$ is uniquely defined \cite{H-SCI}.

Since the continuity bound for the output Holevo quantity presented in Proposition \ref{HQ-C-be-1} does not depend on a
channel, continuity bound for the function $\mu\mapsto \pi_{\mathrm{\Phi}}(\mu)$ under the input average energy constraint can be
obtained by using the same continuity bounds for the functions  $\mu\mapsto \chi(\widehat{\mathrm{\Phi}}(\mu))$  and
$\mu\mapsto \chi(\mathrm{\Phi}(\mu))$. But direct application of Theorem \ref{SCB-1} gives more sharp
continuity bound for the function $\mu\mapsto \pi_{\mathrm{\Phi}}(\mu)$, especially, for degradable and antidegradable channels \cite{DC,Wilde}.

Assume that $H_A$ is the Hamiltonian of system $A$ with the minimal energy $E_0$ satisfying condition (\ref{H-cond+}) and $\hat{F}_{H_A}$ is a function on $\mathbb{R}_+$ satisfying conditions (\ref{F-cond-1}) and  (\ref{F-cond-2}). Denote by $\mathbb{CB}_{\shs t}(\bar{E},\varepsilon\,|\,C,D)$ the expression in the r.h.s. of (\ref{SBC-ineq}).\smallskip

\begin{property}\label{P-C-be-1} \emph{Let $\,\mathrm{\Phi}:A\rightarrow B$ be a quantum channel, $E>E_0$ and $\varepsilon>0$. Then
\begin{equation}\label{P-CB}
\left|\pi_{\mathrm{\Phi}}(\mu)-\pi_{\mathrm{\Phi}}(\nu)\right|\leq \mathbb{CB}_{\shs t}(\bar{E},\varepsilon\,|\,4,2)
\end{equation}
for any ensembles $\shs\mu$ and $\shs\nu$ such that $\,E(\mu),E(\nu)\leq E$ and  $\;D_K(\mu,\nu)\leq\varepsilon$ and any $t\in(0,T]$, where $\bar{E}=E-E_0$ and $T=T(\bar{E},\varepsilon)$ is defined in Theorem \ref{SCB-1}.\footnote{$E(\mu)$ is the average energy of an ensemble $\mu$ defined in (\ref{av-en}).}}

\emph{If $\,\mu\shs$ and $\,\nu\shs$ are discrete ensembles then  the Kantorovich metric $D_K$ can be replaced by any of the metrics  $\,D_0$ and $\,D_*$.}

\emph{If $A$ is the $\ell$-mode quantum oscillator  then (\ref{P-CB}) holds for any $t\in(0,T_*]$ with the r.h.s. replaced by
\begin{equation}\label{P-CB+}
4\varepsilon(1+4t)\left(F_{\ell,\omega}(E)-2\ell\ln(\varepsilon t)+ e^{-\ell}+\ln 2\right)+4g(\varepsilon t)+2g(\varepsilon(1+2t)),
\end{equation}
where $F_{\ell,\omega}$ is the function defined in (\ref{F-ub}) and $\,T_*=(1/\varepsilon)\min\{1, \sqrt{\bar{E}/E_0}\}$.}

\emph{If $\,\mathrm{\Phi}$ is either  degradable or antidegradable channel then (\ref{P-CB}) holds with
$\mathbb{CB}_{\shs t}(\bar{E},\varepsilon\,|\,2,2)$ in the r.h.s. and the first factor $\,4$ in (\ref{P-CB+}) can be replaced by $\,2$.}
\end{property}\smallskip

\textbf{Note:} The r.h.s. of (\ref{P-CB}) does not depend on a channel $\mathrm{\Phi}$. It tends to zero as $\varepsilon\rightarrow0$
for any given $\bar{E}$ and $\shs t \shs $ due to the second condition in (\ref{F-cond-1})  implying uniform continuity of the function
$\mu\mapsto \pi_{\mathrm{\Phi}}(\mu)$ on the set of generalized ensembles with bounded average energy w.r.t. the weak convergence topology.\smallskip

\emph{Proof.} By using Example \ref{privacy} at the end of Section 3.1 and  representations (\ref{St-rep}) and (\ref{c-ch}) it is easy to show
that the function
$$
P_{\mathrm{\Phi}}(\rho)=I(B\!:\!R)_{\mathrm{\Phi}\otimes\mathrm{Id}_{R}(\rho)}-I(E\!:\!R)_{\widehat{\mathrm{\Phi}}\otimes\mathrm{Id}_{R}(\rho)},\quad \rho\in\S(\H_{AR}),
$$
where $R$ is any system, is well defined and satisfies inequality (\ref{F-p-1}) with $a_f=b_f=1$ and  inequality (\ref{F-p-2}) with $c^{-}_f=c^{+}_f=2$
on the set of input states $\rho\in\S(\H_A)$ with finite energy $\Tr H_A\rho$.

Hence Theorem \ref{SCB-1} implies that
\begin{equation}\label{P-CB++}
\left|P_{\mathrm{\Phi}}(\rho)-P_{\mathrm{\Phi}}(\sigma)\right|\leq \mathbb{CB}_{\shs t}(\bar{E},\varepsilon\,|\,4,2)
\end{equation}
for any states $\shs\rho$ and $\shs\sigma$ such that $\Tr H_A\rho,\Tr H_A\sigma\leq E$ and $t\in(0,T]$, where $\bar{E}=E-E_0$.

Continuity bound (\ref{P-CB++}) implies inequality (\ref{P-CB}) for discrete ensembles $\mu$ and $\nu$ such that $\,E(\mu),E(\nu)\leq E$ and  $\;D_*(\mu,\nu)\leq\varepsilon$. It is sufficient to note that $\pi_{\mathrm{\Phi}}(\{p_i,\rho_i\})=P_{\mathrm{\Phi}}(\hat{\rho})$ for arbitrary discrete ensemble $\{p_i,\rho_i\}$, where $\,\hat{\rho}=\sum_i p_i \rho_i\otimes |i\rangle\langle i|\,$ is the qc-state determined by some orthonormal system $\{|i\rangle\}$ in $\H_R$, and to use the arguments from the proof
of Corollary 7 in \cite{AFM}.   It follows from (\ref{d-ineq}) and (\ref{d-ineq+}) that inequality (\ref{P-CB}) holds provided that $\;D(\mu,\nu)\leq\varepsilon$, where $D$ is either $D_0$ or $D_K$.

If $\mu$ and $\nu$ are arbitrary generalized ensembles such that $\,E(\mu),E(\nu)\leq E$ and  $D_K(\mu,\nu)\leq\varepsilon$ then inequality (\ref{P-CB}) can be proved by repeating the arguments from the proof of Proposition 7 in \cite{CID} based on approximation of $\mu$ and $\nu$ by weakly converging sequences of discrete ensembles.

If $\mathrm{\Phi}$ is a degradable (antidegradable) channel then the above function $P_{\mathrm{\Phi}}$ is nonnegative (non-positive). It follows that
(\ref{P-CB++}) holds with $\mathbb{CB}_{\shs t}(\bar{E},\varepsilon\,|\,2,2)$ in the r.h.s.

The assertion  concerning the case when $A$ is the $\ell$-mode quantum oscillator follows from  Corollary \ref{SCB-G-1}. $\square$

\section{Advanced version of the uniform finite-dimensional approximation theorem for capacities of energy-constrained channels}

The uniform finite-dimensional approximation theorem for capacities of energy-constrained channels (the UFA-theorem, in what follows) obtained in \cite{UFA} states, briefly speaking,
that dealing with some  capacity $C_*$  we may assume (accepting arbitrarily small error $\varepsilon$) that \emph{all the channels} have the same finite-dimensional input space -- the subspace corresponding to the $m_{C_*}(\varepsilon)$ minimal eigenvalues of the input Hamiltonian.

The estimates for the  $\varepsilon$-sufficient input dimension $m_{C_*}(\varepsilon)$ obtained in \cite{UFA} for all the basic capacities (excepting $C_{\rm ea}$) turned out extremely hight for small $\varepsilon$ (see Tables 1,2 in \cite{UFA}). In this section we apply the advanced AFW-method
to essentially refine that estimates for $m_{C_*}(\varepsilon)$. This makes the UFA-theorem more applicable for real tasks of quantum information theory.

Assume that $H_A$ is the Hamiltonian of a quantum system $A$ satisfying  condition (\ref{H-cond+}). Then $H_A$ has the representation (\ref{H-rep})
with the orthonormal
basis of eigenvectors $\left\{\tau_k\right\}_{k=0}^{+\infty}$ and the corresponding nondecreasing sequence $\left\{\smash{E_k}\right\}_{k=0}^{+\infty}$ of eigenvalues tending to $+\infty$.

Let $\hat{F}_{H_A}$ be any function on $\mathbb{R}_+$ satisfying conditions (\ref{F-cond-1}) and  (\ref{F-cond-2}), for example, the function $\hat{F}^*_{H_A}$ defined in Proposition \ref{add-l}.
We will use the notations $\,\bar{E}=E-E_0$, $\,\bar{E}_m=E_m-E_0\,$  for all $\,m>0$ and denote by $\mathbb{CB}_{\shs t}(\bar{E},\varepsilon\,|\,C,D)$ the expression in the r.h.s. of (\ref{SBC-ineq}). Let $d_0$ be the minimal natural number such that  $\,\ln d_0>\hat{F}_{H_A}(0)\,$ and $m_0$ a number such that  $\bar{E}_{m}\geq \gamma(d_0)\doteq\hat{F}^{-1}_{H_A}(\ln d_0)$ for all $m\geq m_0$.

A central role in the proof of the original UFA-theorem is played by Lemma 3 in \cite{UFA}. By using the results of Section 3
one can essentially strengthen  this lemma.

\smallskip%\pagebreak
\begin{lemma}\label{b-lemma} \emph{Let $\,\Pi_m(\rho)=P_m\rho P_m+[\Tr(I_A-P_m)\rho]|\tau_0\rangle\langle\tau_0|$, where  $P_m$ is the projector on the subspace $\H^m_A$ corresponding to the minimal $\,m$ eigenvalues $\,E_0,..,E_{m-1}$ of $H_A$ and $\tau_0$ is any eigenvector corresponding to the eigenvalue $E_0$. Let $\,\rho\shs$ be a state  in $\,\S(\H^{\otimes n}_{A}\otimes\H_{R})$ such that $\,\sum_{k=1}^n\Tr H_A\rho_{A_k}\leq nE\,$ and $\,m\geq m_0$. Then
\begin{equation}\label{b-lemma+}
\!\left|I(B^n\!:\!R)_{\Phi^{\otimes n}\otimes\id_{R}(\rho)}-I(B^n\!:\!R)_{\Psi_m^{\otimes n}\otimes\id_{R}(\rho)}\right|\leq n\shs\mathbb{F}_t(u_m, m\,|\shs1),\quad u_m=\sqrt{\bar{E}/\bar{E}_m},
\end{equation}
for any channel $\,\Phi:A\rightarrow B$ and any $\,t\in(0,1]$, where $\,\Psi_m=\Phi\circ\Pi_m$ and
\!\!\begin{equation}\label{f-def}
\begin{array}{rl}
\mathbb{F}_t(u_m, m \,|s)\!\!&\doteq
\displaystyle((4+8t)u_m+2s u_m^2t^2)\hat{F}_{H_{A}}\!\!\left(\bar{E}_m/t^2\right)\\\\ \displaystyle & +\;(4+8t)(1/d_0+\ln 2)u_m
+\;4g(tu_m)
+2g((2+2t)u_m)
\end{array}
\end{equation}
is a quantity tending  to zero as $\,m\rightarrow+\infty$ for any given $\shs t$ and $s\in\{0,1\}$.}\footnote{The function $g(x)$ is defined in (\ref{g-fun}).}
\smallskip

\emph{If  $\,\Tr H_A\rho_{A_k}\leq E\,$ for all $\,k=\overline{1,n}\,$ and $\,\bar{E}\leq\bar{E}_m/t^2\,$ then (\ref{b-lemma+}) holds with $\,n\shs\mathbb{F}_t(u_m, m\,|\shs0)$ in the right hand side. If $\,n=1$ and $\,s_m\doteq \bar{E}/\bar{E}_m+\displaystyle\sqrt{\bar{E}/\bar{E}_m}\leq 2\,$ then   (\ref{b-lemma+}) holds with  $\,n\shs\mathbb{CB}_{\shs t/2}(\bar{E},s_m\,|\,2,2)$ in the right hand side.}
\end{lemma}\medskip

\emph{Proof.} The assumption of the lemma implies that $\,H(\rho_{A_k})<+\infty\,$ for $\,k=\overline{1,n}$.\smallskip

Let $E$ be an environment for the channel $\Phi$, so that the Stinespring representations (\ref{St-rep}) holds with some isometry
$V_{\Phi}$ from $\H_A$ into $\H_{BE}$.

Following  the Leung-Smith telescopic method from \cite{L&S} consider the states
$$
\sigma_k=\Phi^{\otimes k}\otimes\Psi_m^{\otimes (n-k)}\otimes\id_{R}(\rho),\quad k=0,1,...,n.
$$
By repeating the arguments from the proof of Lemma 3  in \cite{UFA} we obtain
\begin{equation}\label{tel}
\left|I(B^n\!:\!R)_{\sigma_n}\!-I(B^n\!:\!R)_{\sigma_0}\right|\leq \displaystyle \sum_{k=1}^n \left|I(B_k\!:\!R|C_k)_{\sigma_k}-I(B_k\!:\!R|C_k)_{\sigma_{k-1}}\right|,
\end{equation}
where $C_k=B^n\setminus B_{k}$ and $I(B_k\!:\!R|C_k)$ is the extended QCMI defined in (\ref{cmi-e+}). The finite entropy of the states $\,\rho_{A_1},...,\rho_{A_n}$, upper bound (\ref{CMI-UB}) and monotonicity of the QCMI under local channels guarantee finiteness of all the terms in (\ref{tel}).

To estimate the $k$-th summand in the r.h.s. of (\ref{tel}) consider  the states
\begin{equation*}%\label{s-one}
\hat{\sigma}_k=V_{\Phi}^{\otimes n}\otimes I_{R} \,\varrho_k\; [V_{\Phi}^{\otimes n}]^*\otimes I_{R}
\end{equation*}
in $\S(\H_{B^nE^nR})$, where $\varrho_k=\id_A^{\otimes k}\otimes\Pi_m^{\otimes (n-k)}\otimes \id_R(\rho)$, $k=0,1,2,...,n$.
The state $\hat{\sigma}_k$ is an extension of the state $\sigma_k$ for each $k$, i.e. $\Tr_{E^n}\hat{\sigma}_k=\sigma_k$. Note that
$[\varrho_k]_{A_j}=\rho_{A_j}$ for $j\leq k$ and $[\varrho_k]_{A_j}=\Pi_m(\rho_{A_j})$ for $j>k$. Hence
\begin{equation}\label{E-est}
  \Tr H_A[\varrho_k]_{A_j}\leq x_j\doteq\Tr H_A\rho_{A_j}\quad\textrm{ for all }k\textrm{ and }j.
\end{equation}

In the  proof of Lemma 3 in \cite{UFA} it is shown that
\begin{equation}\label{norm-est}
\|\hat{\sigma}_k-\hat{\sigma}_{k-1}\|_1\leq2\Tr(I_A-P_m)\rho_{A_k}+2\sqrt{\Tr(I_A-P_m)\rho_{A_k}}\leq 2\varepsilon_k,
\end{equation}
where $\,\varepsilon_k\doteq2\displaystyle\sqrt{\bar{x}_k/\bar{E}_m}$, $\bar{x}_k=x_k-E_0$. \smallskip

Take any $t\in(0,1/2]$. Let $N_1$ be the set of all indexes $k$ for which $\bar{x}_k\leq\bar{E}_m/(4t^2)$ and $N_2=\{1,..,n\}\setminus N_1$. Let $n_i=\sharp (N_i)$, $X_i=\frac{1}{n_i}\sum_{k\in N_i}x_k$ and $\bar{X}_i=X_i-E_0$, $i=1,2$. It follows from (\ref{tel}) that the left hand  side of (\ref{b-lemma+}) do not exceed $S_1+S_2$, where
$$
S_i=\sum_{k\in N_i}|I(B_k\!:\!R|C_k)_{\sigma_k}-
I(B_k\!:\!R|C_k)_{\sigma_{k-1}}|.
$$

For each $k\in N_1$ we have $\varepsilon_k t\leq 1$. Since $\varepsilon^{-1}_k \sqrt{\bar{x}_k/\gamma(d_0)}=\frac{1}{2}\sqrt{\bar{E}_m/\gamma(d_0)}\geq 1/2$ for any $m\geq m_0$, continuity bound (\ref{CB-3}) with (\ref{E-est}) and (\ref{norm-est}) imply (by the arguments used in the proof of Proposition \ref{MI-CB}) that
\begin{equation*}%\label{tel++}
\begin{array}{c}
\displaystyle|I(B_k\!:\!R|C_k)_{\sigma_k}-
I(B_k\!:\!R|C_k)_{\sigma_{k-1}}|\leq \mathbb{CB}_{\shs t}(\bar{x}_k,\varepsilon_k\,|\,2,2)
\\\\
=\displaystyle 4(1+4t)\sqrt{\bar{x}_k/\bar{E}_m}\left(\,\hat{F}_{H_{A}}\!\!\left(\bar{E}_m/(4t^2)\right)+1/d_0+\ln 2\right)\\\\
\displaystyle+4g\!\left(\!2t\sqrt{\bar{x}_k/\bar{E}_m}\right)
+2g\!\left(\!2(1+2t)\sqrt{\bar{x}_k/\bar{E}_m}\right).
\end{array}
\end{equation*}
 Hence, by using the concavity of the functions $\sqrt{x}$ and $g(x)$ along with the monotonicity of $g(x)$ we obtain
\begin{equation}\label{s-1}
\begin{array}{rl}
 S_1\!\! & \displaystyle \leq 4n_1(1+4t)\sqrt{\bar{X}_1/\bar{E}_m}\left(\,\hat{F}_{H_{A}}\!\!\left(\bar{E}_m/(4t^2)\right)+1/d_0+\ln 2\right)\\\\
&\displaystyle+\;4n_1g\!\left(\!2t\sqrt{\bar{X}_1/\bar{E}_m}\right)
+2n_1g\!\left(\!2(1+2t)\sqrt{\bar{X}_1/\bar{E}_m}\right).
\end{array}
\end{equation}

For each $k\in N_2$ the inequality $I(B_k\!:\!R|C_k)\leq I(B_kE_k\!:\!R|C_k)$ and upper bound (\ref{CMI-UB}) imply
\begin{equation*}%\label{tel+++}
\begin{array}{c}
|I(B_k\!:\!R|C_k)_{\sigma_k}-
I(B_k\!:\!R|C_k)_{\sigma_{k-1}}|\leq 2\max\{H([\hat{\sigma}_k]_{B_kE_k}),H([\hat{\sigma}_{k-1}]_{B_kE_k})\}\\\\=2\max\{H([\varrho_k]_{A_k}),H([\varrho_{k-1}]_{A_{k}})\}\leq 2F_{H_{A}}(x_k),
\end{array}
\end{equation*}
where the last inequality follows from (\ref{E-est}). Since $(n-n_2)X_1+n_2X_2\leq nE$ and $X_1\geq E_0$, we have $X_2\leq n\bar{E}/n_2+E_0$. So, by using concavity and monotonicity of the function $\,F_{H_A}$ on $[E_0,+\infty)$ we obtain
\begin{equation}\label{s-2}
\!S_2\leq\sum_{k\in N_2} 2F_{H_{A}}(x_k)\leq 2n_2F_{H_{A}}(X_2)\leq 2n_2F_{H_{A}}(n\bar{E}/n_2+E_0)=2n_2\bar{F}_{H_{A}}(n\bar{E}/n_2).
\end{equation}
It is easy to see that $\bar{X}_1\leq \bar{E}$. Since $\bar{x}_k> \bar{E}_m/(4t^2)$ for all $k\in N_2$ and $(n-n_2)E_0+\sum_{k\in N_2}\bar{x}_k+n_2E_0\leq\sum_{k\in N_1}x_k+\sum_{k\in N_2}x_k\leq nE$, we have $n_2/n\leq 4t^2\bar{E}/\bar{E}_m$. So, it follows from (\ref{s-1}),(\ref{s-2}), concavity of the function $\,\bar{F}_{H_A}$ on $\mathbb{R}_+$ and Lemma 1 in \cite{UFA} that
$$
\begin{array}{c}
\displaystyle\frac{S_1+S_2}{n}\leq 4(1+4t)\sqrt{\bar{E}/\bar{E}_m}\left(\,\hat{F}_{H_{A}}\!\!\left(\bar{E}_m/(4t^2)\right)+1/d_0+\ln 2\right)\\\\
\displaystyle+\;4g\!\left(\!2t\sqrt{\bar{E}/\bar{E}_m}\right)
+2g\!\left(\!2(1+2t)\sqrt{\bar{E}/\bar{E}_m}\right)
+8t^2(\bar{E}/\bar{E}_m)\bar{F}_{H_{A}}\!\!\left(\bar{E}_m/(4t^2)\right).
\end{array}
$$
By replacing $t$ by $t/2$ we obtain the main assertion of the lemma. The vanishing of the quantity $\,\mathbb{F}_t(u_m, m \,|\shs s)\,$ as $\,m\rightarrow+\infty\,$ follows from the second condition in (\ref{F-cond-1}).\smallskip

The assertion concerning the case $\,\Tr H_A\rho_{A_k}\leq E\,$ for all $\,k=\overline{1,n}\,$ follows from the above proof, since in this case the set $N_2$ is empty. In the case $\,n=1$ one can directly apply continuity bound (\ref{CB-3}) with trivial $C$  by using the first inequality in (\ref{norm-est}) with $k=1$, since in this case $\Tr(I_A-P_m)\rho_{A}\leq \bar{E}/\bar{E}_m$ by Lemma 5 in \cite{UFA}. $\square$\smallskip

By using Lemma \ref{b-lemma} one  can obtain an advanced version of the UFA-theorem presented in \cite{UFA}.
In what follows $\,C_*(\Phi ,H_{\!A},E)$, where $C_*$  is one of the capacities $\,C_{\chi}$, $C$, $\bar{Q}$, $Q$, $\bar{C}_{\mathrm{p}}$ and  $C_{\mathrm{p}}$, denotes the corresponding capacity of a quantum channel $\Phi$ from a system $A$ to any system $B$ under the energy constraint determined by the Hamiltonian $H_A$ and energy bound $E$ (see the surveys in Section 4  in \cite{UFA} and in \cite{Wilde+}). In contrast to Theorem 1 in \cite{UFA} we add the non-regularized quantum and private capacities $\bar{Q}$ and $\bar{C}_{\mathrm{p}}$ but exclude the
entanglement-assisted capacity $C_{\mathrm{ea}}$, since for the latter capacity the estimates of the $\varepsilon$-sufficient input dimension obtained in \cite{UFA} are close to tight.

Following \cite{UFA} denote by $\;C_*^m(\Phi ,H_{\!A},E)$ the corresponding capacity of $\Phi$ obtained by block encoding used only states supported by the tensor powers of the $m$-dimensional subspace $\H^m_A$ (defined in Lemma \ref{b-lemma}). It  coincides with the capacity $\;C_*(\Phi_m ,H_{\!A},E)$  of the subchannel $\,\Phi_m$ of $\,\Phi$ corresponding to the subspace $\H_A^m$.

Assume that the Hamiltonian $H_A$ has form (\ref{H-rep}) and $\hat{F}_{H_A}$ is any function on $\mathbb{R}_+$ satisfying conditions (\ref{F-cond-1}) and  (\ref{F-cond-2}).\footnote{By Proposition \ref{add-l} such  function $\hat{F}_{H_A}$ exists if and only if the Hamiltonian $H_A$ satisfies condition (\ref{H-cond+}).} In the following theorem $\mathbb{CB}_{\shs t}(\bar{E},\varepsilon\,|\,2,2)$ is the expression in the r.h.s. of (\ref{SBC-ineq}) with $C=D=2$
and  $\mathbb{F}_t(u_m, m \,|\shs s)$ is the function defined in (\ref{f-def}) for all $m\geq m_0$ and $\mathbb{F}_t(u_m, m \,|\shs s)=+\infty$ otherwise, where $m_0$ is defined before Lemma \ref{b-lemma}. We  use the standard notations $\,\bar{E}=E-E_0$ and $\,\bar{E}_m=E_m-E_0\,$.
\medskip

\begin{theorem}\label{UFA} \emph{Let $C_*$ be one of the capacities $C_{\chi}$, $C$, $\bar{Q}$, $Q$, $\bar{C}_{\mathrm{p}}$ and $C_{\mathrm{p}}$. If the Hamiltonian $H_A$ satisfies condition (\ref{H-cond+}) and $E\geq E_0$ then for any $\,\varepsilon>0$ there exists natural number $\,m_{C_*}(\varepsilon)$  such that
\begin{equation*}%\label{C-approx}
|C_*(\Phi, H_{\!A},E)-C_*^m(\Phi, H_{\!A},E)|\leq\varepsilon\qquad \forall m\geq m_{C_*}(\varepsilon)
\end{equation*}
for arbitrary  channel $\,\Phi$ from the system $A$ to any system $B$.}\smallskip

\emph{The above number $\,m_{C_*}(\varepsilon)$ is the minimal natural number such that $E_m\geq E$ and  $f_{C_*}(E,m,t)\leq\varepsilon$   for at least one $t\in(0,1]$, where}
$$
f_{C_{\chi}}(E,m,t)=f_{\bar{Q}}(E,m,t)=\mathbb{CB}_{\shs t/2}(\bar{E},s_m\,|\,2,2),\quad s_m=\bar{E}/\bar{E}_m+\sqrt{\bar{E}/\bar{E}_m},
$$
$$
f_{C}(E,m,t)=\mathbb{F}_t(u_m, m\,|\shs0),\quad f_{Q}(E,m,t)=\mathbb{F}_t(u_m, m\,|\shs1) ,\quad u_m=\sqrt{\bar{E}/\bar{E}_m},
$$\vspace{5pt}
$$
f_{\bar{C}_p}(E,m,t)=2f_{C_{\chi}}(E,m,t)\quad \textit{and} \quad f_{C_p}(E,m,t)=2f_{Q}(E,m,t).
$$
\smallskip

\emph{If $A$ is the $\ell$-mode quantum oscillator with frequencies $\,\omega_1,...,\omega_{\ell}\,$ then}
\begin{itemize}
\item \emph{the sequence $\{E_k\}_{k\geq0}$  consists of the numbers $\sum_{i=1}^{\ell}\hbar\omega_i(n_i-1/2), n_1,...,n_{\ell}\in \mathbb{N}$ arranged in the nondecreasing order and $\,m_0$ is a number such that $\,E_{m_0}\geq 2E_0$;}
\item \emph{the quantities $\,\mathbb{CB}_{\shs t/2}(\bar{E},s_m\,|\,2,2)$ and $\,\mathbb{F}_t(u_m, m \,|\shs s)$ can be defined, respectively, by the expressions
$$
2s_m(1+2t)\left(\bar{F}_{\ell,\omega}(4\bar{E}/(s_m t)^2)+\Delta^*\right)+2g(s_m(1+t))+4g(s_m t/2),
$$
and
$$
((4+8t)u_m+2s u_m^2t^2)\bar{F}_{\ell,\omega}\left(\bar{E}_m/t^2\right)+(4+8t)u_m \Delta^*
+2g((2+2t)u_m)+4g(tu_m),
$$
where $\bar{F}_{\ell,\omega}(E)$ is the function defined in (\ref{F-ub+}) and $\,\Delta^*=e^{-\ell}+\ln 2$.}
\end{itemize}
\end{theorem}\smallskip

\emph{Proof.} The theorem is proved by repeating the arguments from the proof of Theorem 1 in \cite{UFA}
with the use of Lemma \ref{b-lemma}  instead of Lemma 3 in \cite{UFA}. $\square$
\smallskip

\begin{example}\label{exam}
Let $A$ be the one-mode quantum oscillator with the frequency $\omega$. In this case the Hamiltonian $H_A$ has the spectrum $\{E_k=(k+1/2)\hbar\omega\}_{k\geq 0}$, $F_{H_A}(E)=g(E/\hbar\omega-1/2)$ and $\bar{F}_{1,\omega}(E)=\ln(E/\hbar\omega+1)+1$ \cite[Ch.12]{H-SCI}. The results of numerical calculations of $\,m_{C_*}(\varepsilon)$ for different values of the input energy bound $E$ are presented in the following tables corresponding to two values of the relative error $\varepsilon/F_{H_A}(E)$ equal respectively to $0.1$ and $0.01$.\footnote{All the capacities  $C_*(\Phi, H_{\!A},E)$, $C_*=C_{\chi}, C, Q, \bar{Q}, \bar{C}_{\mathrm{p}}, C_{\mathrm{p}}$, take values in $[\shs0,F_{H_A}(E)\shs]$.}\medskip

 Table 1. The approximate values of $m_{C_*}(\varepsilon)$ for $\varepsilon=0.1F_{H_A}(E)$.

\begin{tabular}{|c|c|c|c|c|c|}
  \hline
  $\;E/\hbar\omega\;$ & $m_{C_{\chi}}(\varepsilon)=m_{\bar{Q}}(\varepsilon)$ & $m_C(\varepsilon)$ & $m_Q(\varepsilon)$ & $m_{\bar{C}_{p}}(\varepsilon)$ & $m_{C_{p}}(\varepsilon)$\\ \hline
  3 & $2.4\cdot10^{5}$ & $9.0\cdot10^{5}$ & $9.0\cdot10^{5}$ & $1.1\cdot10^{6}$ & $4.2\cdot10^{6}$\\ \hline
  10 & $3.7\cdot10^{5}$ & $1.4\cdot10^{6}$ & $1.4\cdot10^{6}$ & $1.6\cdot10^{6}$ & $6.5\cdot10^{6}$\\ \hline
  100 & $1.4\cdot10^{6}$ & $5.3\cdot10^{6}$ & $5.3\cdot10^{6}$ & $6.3\cdot10^{6}$ & $2.5\cdot10^{7}$\\
  \hline
\end{tabular}
\medskip

 Table 2. The approximate values of $m_{C_*}(\varepsilon)$ for $\varepsilon=0.01F_{H_A}(E)$.

\begin{tabular}{|c|c|c|c|c|c|}
  \hline
$\;E/\hbar\omega\;$ & $m_{C_{\chi}}(\varepsilon)=m_{\bar{Q}}(\varepsilon)$ & $m_C(\varepsilon)$ & $m_Q(\varepsilon)$ & $m_{\bar{C}_{p}}(\varepsilon)$ & $m_{C_{p}}(\varepsilon)$\\ \hline
  3 & $3.7\cdot10^{7}$ & $1.5\cdot10^{8}$ & $1.5\cdot10^{8}$ & $1.6\cdot10^{8}$ & $6.5\cdot10^{8}$\\ \hline
  10 & $5.6\cdot10^{7}$ & $2.2\cdot10^{8}$ & $2.2\cdot10^{8}$ & $2.5\cdot10^{8}$ & $1.0\cdot10^{9}$\\ \hline
  100 & $2.1\cdot10^{8}$ & $8.4\cdot10^{8}$ & $8.4\cdot10^{8}$ & $9.5\cdot10^{8}$ & $3.8\cdot10^{9}$\\
  \hline
\end{tabular}
\medskip
\medskip

Comparing the above tables with Tables 1 and 2 in \cite{UFA} shows that Theorem \ref{UFA} gives \emph{substantially smaller} estimates of the $\varepsilon$-sufficient input dimension $m_{C_*}(\varepsilon)$ for all the capacities than Theorem 1 in \cite{UFA}. It is essential that the  estimates of $m_{C_*}(\varepsilon)$ given by
Theorem \ref{UFA} grow with increasing energy (in contrast to the estimates obtained in \cite{UFA}). Since it is clear that real values of $m_{C_*}(\varepsilon)$ must grow with increasing energy, one can assume that the  estimates of $m_{C_*}(\varepsilon)$ given by
Theorem \ref{UFA} are quite adequate. However, the question of the accuracy of these estimates remains open.
\end{example}

\bigskip

{\bf Acknowledgments.}  I am grateful to A.S.Holevo for useful discussion. Many thanks to everyone who asked me about possibility to improve the estimates of the $\varepsilon$-sufficient input dimensions obtained in \cite{UFA} -- these questions  motivated  this work. I am grateful to N.Datta and S.Becker for valuable consultation concerning Theorem 3 in \cite{Datta}.  Special thanks to L.V.Kuzmin for the help with MatLab.

\end{document}